\documentclass[aps,twocolumn,pra,superscriptaddress,amsmath,showpacs,tightenlines]{revtex4}
\usepackage{epsfig,graphicx,times}
\usepackage{amstext}
\usepackage{amsmath}            
\usepackage{amssymb}            
\usepackage{graphicx}           
\usepackage{latexsym}
\usepackage{bm}

\begin{document}

\title{Phonon amplification in two coupled cavities containing one mechanical resonator}

\author{Hui Wang}
\affiliation{Institute of Microelectronics, Tsinghua University,
Beijing 100084, China} \affiliation{Department of Microelectronics
and Nanoelectronics, Tsinghua University, Beijing 100084, China}
\author{Zhixin Wang}
\affiliation{Department of Microelectronics and Nanoelectronics, Tsinghua University,  Beijing 100084, China}

\author{Jing Zhang}
\affiliation{Department of Automation, Tsinghua University, Beijing
100084, P. R. China} \affiliation{Tsinghua National Laboratory for
Information Science and Technology (TNList),
Beijing 100084, China}

\author{\c{S}ahin Kaya \"Ozdemir}
\affiliation{Department of Electrical and Systems Engineering,
Washington University, St. Louis, MO 63130, USA}

\author{Lan Yang}
\affiliation{Department of Electrical and Systems Engineering,
Washington University, St. Louis, MO 63130, USA}

\author{Yu-xi Liu}\email{yuxiliu@mail.tsinghua.edu.cn}
\affiliation{Institute of Microelectronics, Tsinghua University,
Beijing 100084, China}\affiliation{Department of Microelectronics
and Nanoelectronics, Tsinghua University, Beijing 100084, China}
\affiliation{Tsinghua National Laboratory for Information Science
and Technology (TNList), Beijing 100084, China}

\date{\today}

\begin{abstract}
We study a general theory of phonon lasing [I. S. Grudinin \textit{et al.},
Phys. Rev. Lett. {\bf 104}, 083901 (2010)] in coupled
optomechancial systems. We derive the dynamical equation of the
phonon lasing using supermodes formed by two cavity modes. A general threshold condition
for phonon lasing is obtained. We also show the differences between phonon lasing and photon lasing , generated by photonic supermodes and two-level atomic systems, respectively. We find that the phonon lasing can
be realized in certain parameter regime near the threshold. The phase diagram and second-order correlation function
of the phonon lasing are also studied to show
some interesting phenomena that cannot be observed in the common
photon lasing with the two-level systems.

\end{abstract}

\maketitle \pagenumbering{arabic}

\section{Introduction}

Phonons are quanta of sound and are very important concept in
condensed matter physics. They exist as vibrating modes  in various
physical systems, including single trapped ions~\cite{iontrap}, atoms in
solid-state materials~\cite{Ashcroft},  and macroscopic mechanical
resonator~\cite{physicsrep}. Phonons are very similar to photons
in the sense that both are bosonic particles and obey
Bose-Einstein statistics. Thus, the phenomena occurred in photons
are quite often brought to phonons. For instance, in analogue to
coherent and squeezed photon states, squeezed phonon
states~\cite{xuedong} have been proposed to be generated in bulk
solid-state materials by using second-order Raman scattering and
explored to modulate quantum fluctuations of atomic displacements.
The coherent phonon generation via impulsive stimulated Raman
scattering in the condensed media~\cite{coherent} and the phonon
stimulated emission have also been
studied~\cite{PT1,PT2,PT3,Bahl}.

Reaching the  quantum mechanical regime for vibrating modes of
macroscopic mechanical resonators is a longstanding
goal~\cite{physicsrep}. Both mechanical resonators and single-mode
cavity fields can be modeled as harmonic oscillators. Thus, it is
reasonable to expect that mechanical resonators can play the
role of single-mode cavities in quantum optics, and help to realize
mechanical quantum electrodynamics (QED) by replacing the
single-mode cavity field with a mechanical
resonator in its quantum regime. Experimentalists have showed that the usual Jaynes-Cumming
model in the cavity QED can be realized by coupling a superconducting qubit to a mechanical
resonator~\cite{OConnell,Lahaye1,Lahaye2}. Also several methods
have been developed in optomechanical systems for cooling the mechanical resonators with
low frequencies to their quantum ground states by coupling them to single-mode microwave or
optical fields via the radiation pressure~\cite{Vahalareview}. These achievements lay
a solid foundation to develop single-mode phonon
cavity~\cite{Mahboob-naturephyscs} and manipulate phonon states~\cite{xunwei1,xunwei2} at single-photon level~\cite{Jieqiao1,Jieqiao2}
using mechanical resonators.

It well known that lasing can be generated when the stimulated
emission is coherently amplified by a gain medium inside a
cavity. Similar to the lasing, several theoretical proposals were
put forward to generate phonon coherent amplification of
stimulated emission in different systems, e.g., the quantum
dot~\cite{Kabuss}, ultra-cold atomic gas~\cite{Mendonca},
nanomagnets~\cite{Garanin}, acoustical cavities~\cite{Fokker}, and
double barrier systems~\cite{Makler1}. The amplification of
mechanical oscillations was also theoretically studied by coupling
a nanomechanical resonator to polarized paramagnetic
nuclei~\cite{Bargatin}. Such so-called phonon lasing was
experimentally demonstrated in the systems of vibrating
microscopic particles (e.g.,trapped magnesium ions~\cite{vahala-nature}),
electromechanical resonator~\cite{Yamaguchi}, and superlattice
structures~\cite{Beardsley}. Recently reported several experiments on
phonon lasing in optomechanical systems~\cite{Khurgin1,Khurgin2}
and also in an optomechanical system coupled to a cavity~\cite{Vahala2}
are of particular interest, and have been focus of attention.

In the experiment reported in Ref.~\cite{Vahala2}, two degenerate microtoroid
whispering gallery  mode (WGM) optical resonators are coupled in a controllable
way through the evanescent fields, and one of them is coupled to a
mechanical resonator via radiation pressure. These two coupled
optical cavities form two supermodes acting as a controllable
two-level system which can be tuned to resonantly interact with
the mechanical resonator. This ingenious design is very similar to
the laser system with two-level atoms interacting with single-mode
cavity fields~\cite{scully}. Although the threshold condition on
phonon lasing in the coupled cavity system has been
discussed in Ref.~\cite{Vahala2}, the stability of the system,
 lasing phonon statistics and the threshold still need to
be rigorously analyzed. Motivated by the experiment on phonon
lasing~\cite{Vahala2} and considering both
theoretical~\cite{gong,haixing,Basiri-Esfahani,Marquardt1,Marquardt3,Rabl2,xu}
and experimental~\cite{Massel,zhang} studies on the coupled
optomechanical systems from optical to microwave
frequencies~\cite{Vahalareview,Regal}, we have analyzed phonon
lasing in coupled optomechanical systems by using a completely quantum theoretical approach.

\begin{figure}
\includegraphics[bb=127 178 350 410, width=8 cm, clip]{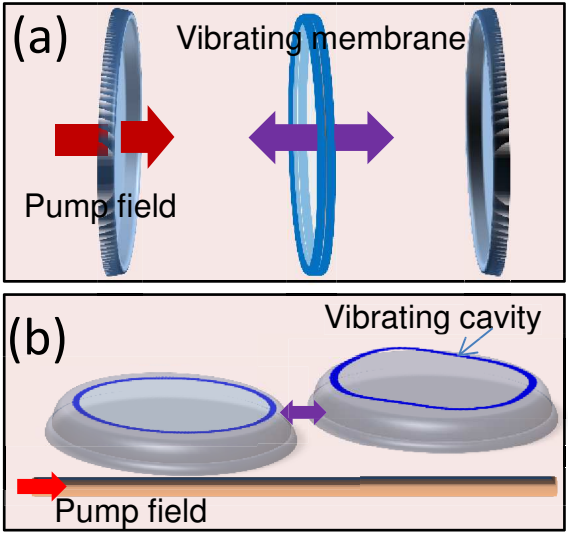}
\caption[]{(Color online) (a) A movable mirror is placed inside a
cavity, and is coupled to the optical modes $a_{L} $ and  $a_{R} $ of
the left and right cavity. (b) A cavity is coupled to an optomechanical system.
In both cases, we assume that the two cavity fields have the same
 frequencies and  decay rates. A classical driving field
is applied to the left cavity.}\label{fig1}
\end{figure}

Our paper is organized as follows. In Sec.~II, we give the
Hamiltonian which is used to describe possible experimental setups,
and then obtain the supermode description for the phonon lasing. In
Sec.~III, we derive the steady state solution using the equations of
motion for the variables of the cavity fields and the mechanical
resonator. We also study how the phonon number, population inversion
of the supermodes, and pump power change with the driving field. In
Sec.~IV, we carefully study the phase diagram of the phonon lasing
by using equations of motion in the phonon stimulated regime. In
Sec.~V, the phonon statistics is studied. Finally,
 we summarize and discuss our results.

\section{Theoretical model}

As schematically shown in Fig.~\ref{fig1}, we study two very
similar systems: The first one contains a vibrating dielectric
membrane (e.g., in Ref.~\cite{Thompson}), which is placed midway
 between two mirrors as in Fig.~\ref{fig1}(a), the
Hamiltonian of the system is described in
Refs.~~\cite{gong,haixing,Basiri-Esfahani,Marquardt1}; the second
one consists of a cavity, which is coupled to an optomechanical
system (e.g., in Ref.~\cite{Vahala2}) as in Fig.~\ref{fig1}(b),
the Hamiltonian of the system is described as in
Refs.~\cite{Marquardt3,Rabl2,xu}. The difference between these systems is as follows.
 In the first system, both the left and the right cavity fields interact
with the mechanical resonator, whereas in the second system only
the right cavity field interacts with the mechanical resonator.
 Besides a factor of $1/2$ difference in the
coupling strength between the supermodes and the mechanical
resonator described below in Eq.~(\ref{eq:1}), we find that two
systems have the same Hamiltonian in the presentation of the
supermodes. Thus we only deal with the Hamiltonian corresponding to Fig.~\ref{fig1} (a)
in the following calculations.

We assume that the classical pump field with the frequency
$\omega_{d}$ is applied to the left cavity and the two cavities have
the same frequencies $\omega_{c}$ when the mechanical resonator is
in its equilibrium position.  We define the annihilation (creation)
operators for the left and the right cavity fields as $a_{L}$
($a^{\dagger}_{L}$) and $a_{R}$ ($a^{\dagger}_{R}$). By using the
supermode operators $a_{1}=(a_{L}+a_{R})/\sqrt{2} $ and
$a_{2}=(a_{L}-a_{R})/\sqrt{2}$ as in Ref.~\cite{Vahala2}, the
Hamiltonian $H_{0}$ for both systems driven by the external field
can be written as
\begin{eqnarray}\label{eq:1}
H_{0}&=&\hbar\omega_{m} b^{\dagger}b-\hbar\chi\left(a^{\dagger}_{1}a_{2}b
+a^{\dagger}_{2}a_{1}b^{\dagger}\right)+\hbar \left(g-\Delta\right)a^{\dagger}_{1}a_{1}\nonumber\\
&-&\hbar\left(g+\Delta\right)a^{\dagger}_{2}a_{2}+\frac{i\hbar}{\sqrt{2}}\left[\Omega(a^{\dagger}_{1}
+a^{\dagger}_{2})-\text{h.c.}\right],
\end{eqnarray}
with the frequencies $g-\Delta$ and $g+\Delta$ for supermodes $1$
and $2$. Here, $\chi$ represents the coupling strength between the
cavity field and the mechanical resonator via the radiation
pressure. We note that $\chi$ should be changed into $\chi/2$ for
the system shown in Fig.~\ref{fig1}(b). $\Delta=\omega_{d}-\omega_{c}$ is the detuning
between the diving field and cavity mode. $b$ and $b^{\dagger}$
denote the annihilation and creation operators of the mechanical
mode with the frequency $\omega_{m}$. The interaction strength
between the two cavities is described by the parameter $g$. The
coupling constant between the driving field and the left cavity
field is $\Omega$.

Using the Hamiltonian in Eq.~(\ref{eq:1}) and also considering the
environmental effect, we can obtain equations of motion for all
variables of the system as
\begin{eqnarray}
\frac{d
a_{1}}{dt}&=&-\left[\frac{\gamma_{c}}{2}+i(g-\Delta)\right]a_{1}
+i\chi a_{2}b+\frac{\Omega}{\sqrt{2}}+\Gamma_{1}(t), \label{eq:3}\\
\frac{d
a_{2}}{dt}&=&-\left[\frac{\gamma_{c}}{2}-i(g+\Delta)\right]a_{2}
+i\chi a_{1}b^{\dagger}+\frac{\Omega}{\sqrt{2}}+\Gamma_{2}(t), \label{eq:4}\\
\frac{d b}{dt}&=&-\left(\gamma_{m} +i\omega_{m}\right) b+i\chi
a^{\dagger}_{2}a_{1}+\sqrt{2\gamma_{m}}b_{\rm{in}}.\label{eq:5}
\end{eqnarray}
Here, we assume that decay rates $\gamma_{L}$ and $\gamma_{R}$ of
the left and right cavities are same, i.e.,
$\gamma_{L}\equiv\gamma_{R}=\gamma_{c}$, and the decay rate of the mechanical
resonator is assumed as $\gamma_{m}$. $\Gamma_{1}(t) $,
$\Gamma_{2}(t) $, and $b_{\rm in}(t) $ represent fluctuation operators
corresponding to the supermodes and the mechanical resonator. As shown
in Appendix~\ref{A1}, the correlation functions of the fluctuation
operators of the supermodes in the time domain under the Markovian
approximation are given as
\begin{eqnarray}
&\langle \Gamma_{1}(t) \Gamma^{\dagger}_{1}(t^{\prime})\rangle=\langle \Gamma_{2}(t) \Gamma^{\dagger}_{2}(t^{\prime})\rangle=\gamma_{c}\delta(t-t^{\prime}),\label{eq:n1}&\\
&\langle \Gamma^{\dagger}_{1}(t)
\Gamma_{1}(t^{\prime})\rangle=\langle \Gamma^{\dagger}_{2}(t)
\Gamma_{2}(t^{\prime})\rangle=0.\label{eq:n2}&
\end{eqnarray}
Here, we assume that the thermal energy due to environmental
temperature $T$ is sufficiently lower than the transition energies
of the supermodes, thus the effect of the environmental
temperature on two supermodes is negligibly small. We assume that the frequency of the mechanical resonator
is low, thus the temperature effect on the mechanical resonator should be included. That is, the correlation
functions of the fluctuation operators of the mechanical resonator
is assumed as
\begin{eqnarray}
&\langle b^{\dagger}_{\rm {in}}(t)b_{\rm {in}}(t^{\prime})\rangle= n_{b}(T) \delta(t-t^{\prime}),&  \label{eq:n3}\\
&\langle b_{\rm {in}}(t^{\prime})b^{\dagger}_{\rm {in}}(t)\rangle
= [n_{b}(T)+1] \delta(t-t^{\prime}),\label{eq:n4}&
\end{eqnarray}
where $n_{b}(T)=1/[\exp\left(\hbar\omega_m/K_B T\right)-1]$
denotes the thermal phonon number
of the mechanical resonator, with the Boltzmann constant $K_{B}$.

\section{Steady states analysis}

In this section, we analyze the steady-state of the
coupled optomechanical systems. First, we show that the
bistable behavior can be observed in these coupled systems even in the blue detuning regime. Then, we
analyze the stability under
small disturbances in the parameter regions of phonon coherent
amplification.

\subsection{Mechanical bistability}

\begin{figure}
\includegraphics[bb=40 210 450 630, width=7 cm, clip]{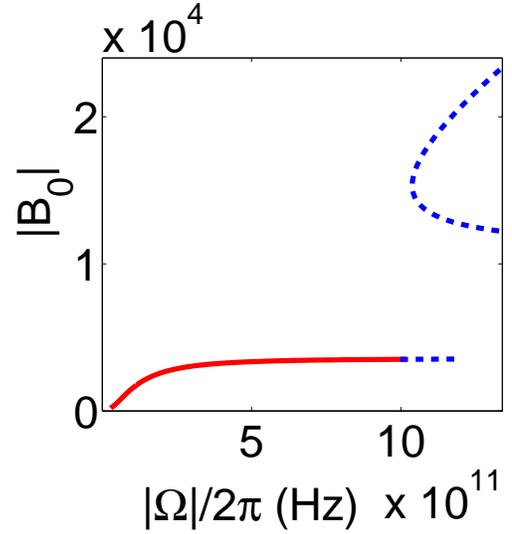}
\caption[]{(Color online) The steady-state value $|B_{0}|$ is plotted
as function of the strength $|\Omega|/(2\pi)$ of driving field.
The other parameters are assumed as
$\omega_m/(2\pi)=23.4$ MHz, $\chi/(2\pi)=1570$ Hz, $\gamma_{m}/(2\pi)=0.125$ MHz,
 $\gamma_c/(2\pi)=4.8$ MHz, $g/(2\pi)=11.7$ MHz, and $\Delta=g/2$.
}\label{fig2}
\end{figure}

By using the semiclassical approximation, e.g., $\langle a_{2} b\rangle =\langle a_{2}\rangle \langle b\rangle$, we obtain
the average value of the Eqs.(2)-(4) as
\begin{eqnarray}
\langle \dot{a}_{1}\rangle&=&-\left[\frac{\gamma_{c}}{2}+i(g-\Delta)\right]\langle a_{1}\rangle+i\chi\langle a_{2}\rangle \langle  b\rangle+\frac{\Omega}{\sqrt{2}},  \label{eq:6}\\
\langle \dot{a}_{2}\rangle&=& -\left[\frac{\gamma_{c}}{2}-i(g+\Delta)\right]\langle a_{2}\rangle+i\chi \langle a_{1}\rangle\langle b^{\dagger}\rangle+\frac{\Omega}{\sqrt{2}}, \label{eq:7}\quad\\
\langle\dot{ b}\rangle&=&-\gamma_{m} \langle b\rangle-i\omega_{m}\langle b\rangle+i\chi \langle a^{\dagger}_{2}\rangle \langle a_{1}\rangle.\label{eq:8}
\end{eqnarray}
Defining the steady state values of the cavity fields and the mechanical resonator as
$\langle a_{1}\rangle_{s}=A_{1}$,  $\langle a_{2}\rangle_{s}=A_{2}$,  and
 $\langle b\rangle_{s}=B_{0}$ respectively, and then setting
$\dot{\langle a_{1}\rangle_{s}}=\dot{\langle a_{2}\rangle_{s}}=\dot{\langle b\rangle_{s}}=0$,
 we obtain the following relation between the driving
strength $\Omega$ and the mechanical mode $B_0$
\begin{equation}
\left|\Omega\right|^2=\frac{2B_0\left(\gamma_{m}
+i\omega_{m}\right)\left[\left(|\xi|^{2}-\Delta^{2}+\chi^{2}
|B_{0}|^{2}\right)^{2}+\gamma^{2}_{c}\Delta^{2}\right]}{i\chi\left[\xi^{2}+\left(\chi
B_{0}-\Delta\right)^{2}\right]}.\label{eq:9}
\end{equation}
where we have defined $\xi=\gamma_{c}/2-ig$.
As shown in Fig.~\ref{fig2}, the red solid curve represents the
stable region of phonon mode, while blue dashed curve represents the
boundary between stable and unstable regions. When the strength
of driving field is increased to a critical point, the system becomes
unstable and the bistability behaviors of phonon mode can be
observed. In contrast to the results of single-mode optomechanical
system where bistability can be shown only in the red-detuning
regime~\cite{Aldana}, the bistability of coupled  optomechanical
systems can be observed in the blue-detuning regime.

From Fig.~\ref{fig2}, we see that the curve of the steady-state value
$|B_{0}|$ of the phonon mode versus the strength of the driving
field $|\Omega|$ is not in the usual S-shape. This is because the driving field is now
in the strong regime which leads to so called unconventional
bistability behaviors~\cite{Liew}.

\subsection{Stabilities}
To discuss the stability of the system against small
perturbations, let us now express the variables $a_1$, $a_2$, and
$b$ as the sums of their stable steady-state values and small
fluctuations, that is,
\begin{eqnarray}
a_{1}(t)&=&A_1+\Lambda_{1}(t),\label{eq:10}\\
a_{2}(t)&=&A_2+\Lambda_{2}(t), \label{eq:11}\\
b(t)&=&B_{0}+\beta(t).\label{eq:12}
\end{eqnarray}
The expressions of $a^{\dagger}_{1}(t)$, $a^{\dagger}_{2}(t)$ and
$b^{\dagger}(t)$ can be obtained from the Hermitian conjugates of
$a_{1}(t)$, $a_{2}(t)$ and $b(t)$ correspondingly. Here, the
average values of fluctuation terms are zero, i.e.,
$\langle\Lambda^{\dagger}_{1}(t)\rangle=\langle\Lambda^{\dagger}_{2}(t)\rangle=\langle\beta(t)\rangle=0$.
The steady-state values $A_{1}$, $A_{2}$, $B_{0}$ and their
complex conjugates can be easily obtained from Eqs.~(\ref{eq:6})-
(\ref{eq:8}) by setting all time derivatives to zero. Then we can write down the linearized dynamical
equations of the fluctuation terms up to the first order as
\begin{eqnarray}\label{eq:13}
\frac{d}{d t}\vec{u}=M \vec{u}.
\end{eqnarray}
Here, the matrix $M$ is given as
\begin{equation}
M =i\chi\left(
\begin{array}{cccccc}
\frac{\epsilon_{1}}{i\chi} & 0 &  B_{0} & 0 &  A_{2} & 0 \\
0 &\frac{\epsilon^{\ast}_{1}}{i\chi}  & 0 & -B^{\ast}_{0} & 0 & - A^{\ast}_{2} \\
 B^{\ast}_{0} & 0 &  \frac{\epsilon_{2}}{i\chi} & 0 & 0 &  A_{1} \\
0 & -B_{0} & 0 & \frac{\epsilon^{\ast}_{2}}{i\chi} &  - A^{\ast}_{1} & 0 \\
 A^{\ast}_{2} & 0 & 0 &  A_{1} & \frac{\epsilon_{3}}{i\chi} & 0 \\
0 & -A_{2}  & - A^{\ast}_{1} &0 & 0 & \frac{\epsilon^{\ast}_{3}}{i\chi} \\
\end{array}
\right)\label{eq:17-1}
\end{equation}
with $\epsilon_{1}=-(\gamma_{c}/2)-i(g-\Delta)$,
$\epsilon_{2}=-(\gamma_{c}/2)+i(g+\Delta)$, and
$\epsilon_{3}=-\gamma_{m}-i\omega_{m} $. The vector $\vec{u}$ in
Eq.~(\ref{eq:13}) is defined as $\vec{u}=(\Lambda_{1}\,\,
\Lambda^{\dagger}_{1}\,\, \Lambda_{2}\,\, \Lambda^{\dagger}_{2}\,\,
\beta\,\, \beta^{\dagger} )^{T}$ where the superscript $T$ denotes
the transpose.

\begin{figure}
\includegraphics[bb=-91 171 510 690, width=4.2 cm, clip]{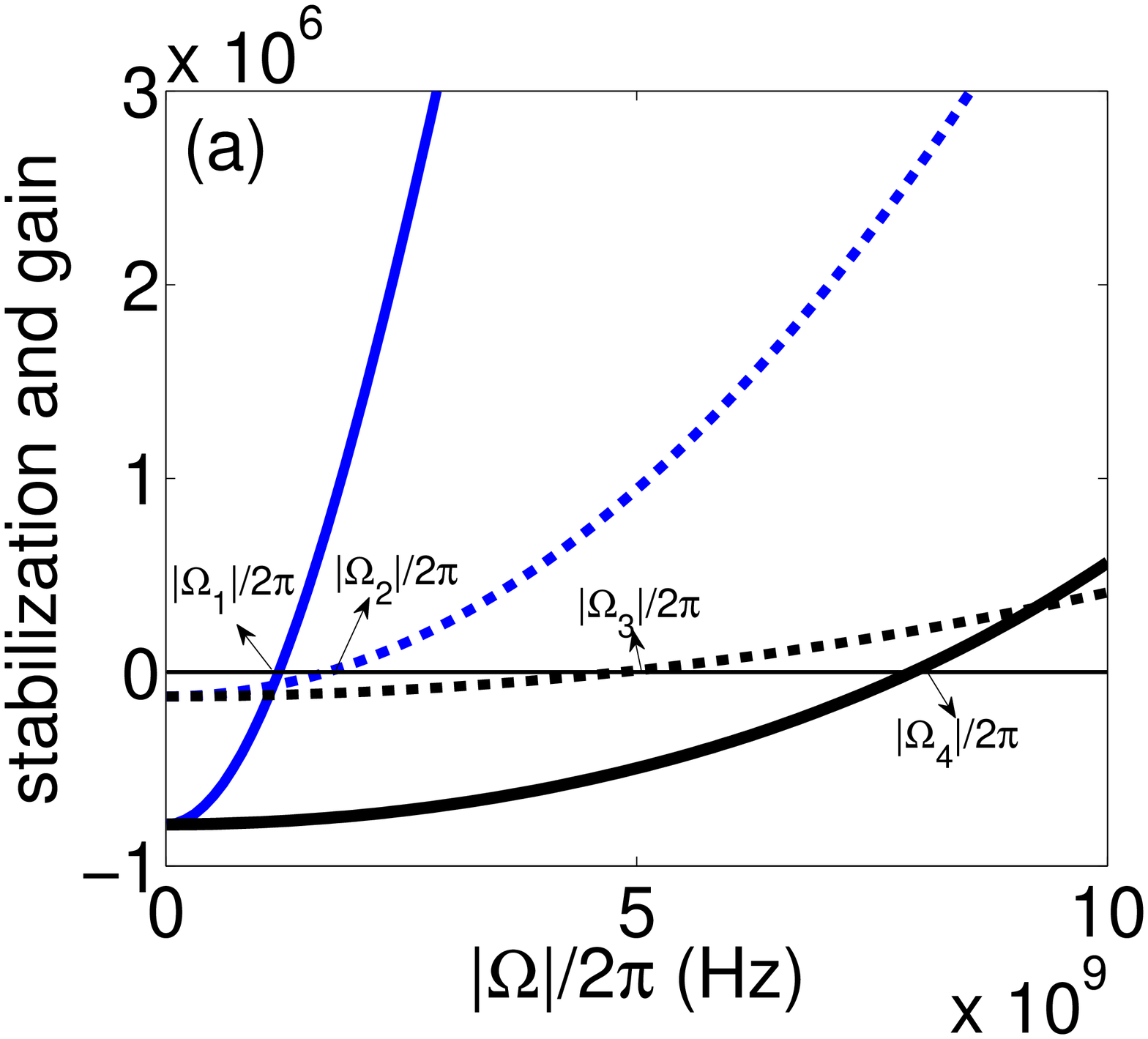}
\includegraphics[bb=-91 171 510 690, width=4.2 cm, clip]{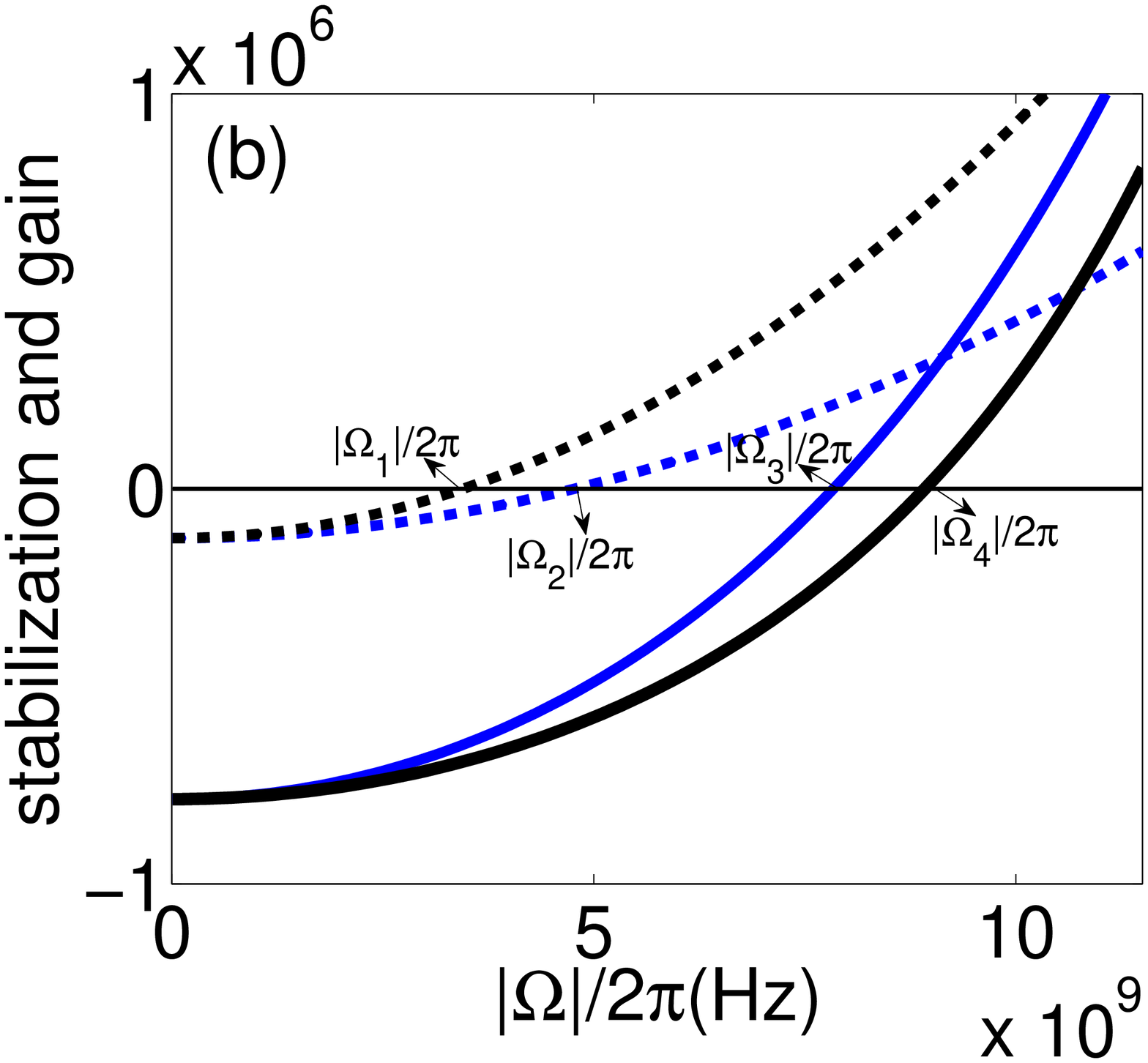}
\caption[]{(Color online) The real part of the eigenvalues
$\text{Re}{(\lambda)}$ of the matrix $M$ in Eq.~(\ref{eq:17-1}) and the net gain $\alpha^{\prime}$
in Eq.~(\ref{eq:25}) of phonons versus the strength $|\Omega|$ of the driving field. Solid
curves describe $\text{Re}{(\lambda)}$, and the dashed curves
correspond to $\alpha^{\prime}$. (a) Blue
curves correspond to $\Delta=g$, while the black curves correspond to $\Delta=g/2$.
 We set  $\gamma_c/(2\pi)=4.8$ MHz. The points $|\Omega_{1}|/(2\pi)$ and
$|\Omega_{4}|/(2\pi)$ correspond to ${\rm Re}{(\lambda)}=0$, however $|\Omega_{2}|/(2\pi)$ and
$|\Omega_{3}|/(2\pi)$  correspond to $\alpha^{\prime}=0$.
(b) The blue  curves
correspond to $\gamma_{c}/(2\pi)=4.8$ MHz, while the black curves
correspond to $\gamma_{c}/(2\pi)=2.8$ MHz. Here, we set $\Delta=g/2$.
The points $|\Omega_{3}|/(2\pi)$ and
$|\Omega_{4}|/(2\pi)$ correspond to ${\rm Re}{(\lambda)}=0$, whereas $|\Omega_{1}|/(2\pi)$ and
$|\Omega_{2}|/(2\pi)$  correspond to $\alpha^{\prime}=0$.
The other parameters are the same as in
Fig.~\ref{fig2} except coupling strength $g$ between two cavities, its value can be obtained from the quantity
 $\delta/(2\pi)=(2g-\omega_{m})/(2\pi)=1$ MHz.
}\label{fig3}
\end{figure}

The stability of the system is determined by the eigenvalues of
the matrix $M$. If the real parts of the eigenvalues of $M$ are
all negative, the system is stable. Otherwise the system is
unstable. It is not easy to analytically solve the eigenvalues of
$M$. However, we can numerically calculate them. The matrix $M$
includes three pairs of conjugate eigenvalues, we find that the
real parts of two pairs of eigenvalues are always negative near
the threshold, thus we  need only to calculate the real part
 ${\rm Re}{(\lambda)}$ of the eigenvalues for the remaining pair.
 In Fig.~\ref{fig3}, the
real parts ${\rm Re}{(\lambda)}$ of these two eigenvalues as a function
 of the strength $|\Omega|$ of the driving field is plotted for given
 parameters.  From Fig.~\ref{fig3}(a), we find that the system has a
 larger stable region, i.e., ${\rm Re}{(\lambda)}<0$, for the strength
 $|\Omega|$ of the driving field in the case  $\Delta=g/2$ in contrast to
 $\Delta=g$.  Figure~\ref{fig3}(b) shows that the stable region of the system
  becomes smaller for faster decay rates of the cavity field. As a comparison
  with the stability of the system,  the gain of the phonon lasing is also
  plotted as a function of the strength $|\Omega|$ of the driving field in Fig.~\ref{fig3}.
The detailed discussion will be given when Eq.~(\ref{eq:25}) is introduced.

\section{Phonon lasing}

Let us now discuss the important properties of the phonon
lasing. Under the adiabatic approximation of angular
operators of supermodes, we derive the dynamical equation of the
phonon mode and revisit the threshold condition of phonon lasing which was
obtained in Ref.~\cite{Vahala2}. We also define an effective
potential for the phonon mode near the threshold in a special
case, and discuss the phase diagram of phonon lasing.

\subsection{Phonon lasing equation and threshold condition}

To find the relation between the phonon lasing and
population inversion of two supermodes, and also to compare
 the phonon lasing with the photon lasing
generated by two-level atomic system, we define the ladder and
population inversion operators via the angular momentum operators
constructed by two bosonic supermode operators~\cite{sakura}.
Because the energy of the supermode $1$ is larger than that of the
supermode $ 2 $, then we define the angular momentum operators as:
$J_{+}=a^{\dagger}_{1}a_{2}$, $J_{-}=a^{\dagger}_{2}a_{1}$ and
$J_{z}=(a^{\dagger}_{1}a_{1}-a^{\dagger}_{2}a_{2})/2$. It is clear that
$J_{+}$ and $J_{-}$ are ladder operators, while $J_{z}$ describes the population inversion from
the supermode mode $1$ to $2$.

Using Eqs.~(\ref{eq:3})-(\ref{eq:5}) and also their Hermitian
conjugates, the dynamical equations for the
variables $J_{-}$ and $J_{z}$ can be obtained as below
\begin{eqnarray}
\frac{d J_{-}}{d t}&=&-\gamma_{c}J_{-}-2igJ_{-}-2i\chi J_{z}b+\frac{\Omega^{\ast}}{\sqrt{2}}a_{1}+\frac{\Omega}{\sqrt{2}}a^{\dagger}_{2}\nonumber\\
&+&\Gamma^{\dagger}_{2}(t)a_{1}+a^{\dagger}_{2}\Gamma_{1}(t),\label{eq:14}\\
\frac{d J_{z}}{d t}&=&-\gamma_{c}J_{z}+i\chi J_{+}b-i\chi J_{-}b^{\dagger}+\frac{\Omega}{2\sqrt{2}}a^{\dagger}_{1}+\frac{\Omega^{\ast}}{2\sqrt{2}}a_{1}\nonumber\\
&-&\frac{\Omega}{2\sqrt{2}}a^{\dagger}_{2}-\frac{\Omega^{\ast}}{2\sqrt{2}}a_{2}
+\frac{1}{2}a^{\dagger}_{1}\Gamma_{1}(t)+\frac{1}{2}\Gamma^{\dagger}_{1}(t)a_{1}\nonumber\\
&-&\frac{1}{2}a^{\dagger}_{2}\Gamma_{2}(t)-\frac{1}{2}\Gamma^{\dagger}_{2}(t)a_{2},\label{eq:15}\\
\frac{d b}{d t}&=&-\gamma_{m} b-i\omega_{m} b+i\chi
J_{-}+\sqrt{2\gamma_{m}}b_{\rm in}(t).\label{eq:16}
\end{eqnarray}
Here, we note that unpaired operators of the optical supermodes
are still remained in Eqs.~(\ref{eq:14}) and (\ref{eq:15}).
Eqs~(\ref{eq:14})-(\ref{eq:16}) for the optical
supermodes interacting with the phonon mode are very similar to
those of photon lasing~\cite{Haken}. However, we can also easily find the difference
between the phonon lasing, described in Eqs. (\ref{eq:14}) and (\ref{eq:15}),
and the photon lasing in a two-level system~\cite{Haken}. Since additional terms $(\Omega^*a_{1}+\Omega
a_{2}^{\dagger})/\sqrt{2}$ appear in Eq.~(\ref{eq:14}) and
$[\Omega^{*}(a_{1}- a_{2})+\text{h.c.}]/2\sqrt{2}$ in
Eq.~(\ref{eq:15}), the phonon lasing studied here will show different behaviors
compared to the photon lasing with two-level systems~\cite{Haken}.

We assume that the decay rate $\gamma_{c}$ of the cavity field is
much larger than the decay rate $\gamma_{m}$ of the mechanical
resonator, then the variables $ J_{-}$ and $ J_{z}$ of the cavity
field will be subject to the dynamics of the mechanical resonator
under the conditions that $\partial J_{-}/(\gamma_{c}
\partial t)\ll J_{-}$ and $\partial J_{z}/(\gamma_{c}
\partial t)\ll J_{z}$ which leads to the so-called adiabatical approximation.
By setting $\partial J_{-}/\partial t=0$ and $\partial
J_{z}/\partial t=0$, we can obtain the expression of $J_{-}$ and
$J_{z}$. Substituting the expression of $J_{-}$ into the dynamical
equation of $b$ in Eq.~(\ref{eq:16}), we have
\begin{eqnarray}
\frac{d b}{d t}&=&-i\omega_{m}b-\left[\gamma_{m}-2\frac{\chi^2
J_{z}}{\gamma_{c}+i\delta}\right]b
+\frac{\alpha}{\sqrt{2}}\left(\Omega^{\ast}a_{1}+\Omega
a^{\dagger}_{2}\right)\nonumber \\
&+&\alpha\left[\Gamma^{\dagger}_{2}(t)a_{1}
+a^{\dagger}_{2}\Gamma_{1}(t)\right]+\sqrt{2\gamma_{m}}b_{in}(t),\label{eq:20}
\end{eqnarray}
where $\alpha=i\chi/(\gamma_{c}+i\delta)$ and
$\delta=2g-\omega_{m}$ is a small detuning between two-level
system formed by supermodes and the frequency of the mechanical resonator (see the
fast oscillating term $i2gJ_{-}$ in Eq.~(\ref{eq:14})).  From Eq.~(\ref{eq:20}), we find the phonon lasing gain as
\begin{equation}\label{eq:21}
G^{\prime}={\rm Re}\left(\frac{2\chi^2
J_{z}}{\gamma_{c}+i\delta}\right)=\frac{2\chi^2\gamma_{c}J_{z}}{\gamma^2_{c}+(2g-\omega_{m})^2}
\end{equation}
if we omit the single mode terms related to $a_1$ and $a_2$ in Eq.~(\ref{eq:20}).
We note that this is just the phonon lasing gain obtained in
Ref.~\cite{Vahala2}. The gain
described by Eq.~(\ref{eq:21}) is not so simple when the third term
of the right hand of Eq.~(\ref{eq:20}) is taken into account.

To obtain more exact expression for the gain of
phonon lasing , let us further adiabatically eliminate $a_{1}$
and $a_{2}$ in Eq.~(\ref{eq:20}) by setting $\partial
a_{i}/\partial t =0$ (with $i=1,2$) in Eqs.~(\ref{eq:3}) and
(\ref{eq:4}). In this case, equation~(\ref{eq:20}) can be
reexpressed as
\begin{eqnarray}\label{eq:22}
\frac{d b}{d t}&=&-\left[\gamma_{m}-\frac{2\chi^2
J_{z}}{\gamma_{c}+i\delta}
+\frac{i\Delta\gamma_{c}\chi^{2}|\Omega|^{2}}{\left(\gamma_{c}+i\delta\right)
(N^{2}+\Delta^2\gamma^2_{c})}\right]b  \nonumber\\
&-&i\omega_{m}b+i\frac{\chi|\Omega|^{2}\left[N(\gamma_{c}-i2g)+2\Delta^2\gamma_{c}\right]}{2(\gamma_{c}+i\delta)
(N^2+\Delta^2\gamma^2_{c})} +\Gamma(t),\qquad
\end{eqnarray}
where the parameter $N$ is defined as
$N=\gamma^{2}_{c}/4+g^2-\Delta^2+\chi^2 b^{\dagger}b$; and the
noise term $\Gamma\left(t\right)$ can be found in Eq.~(\ref{eq:A9-1}) in Appendix~\ref{A1}.
Thus, the gain $G$ of the phonon lasing from
Eq.~(\ref{eq:22}) becomes
\begin{equation}\label{eq:25-1}
G=G^{\prime}-\frac{\Delta\delta\gamma_{c}\chi^{2}|\Omega|^{2}}
{\left(\gamma^2_{c}+\delta^2\right) (N^{2}+\Delta^2\gamma^2_{c})}
\end{equation}
where $G^{\prime}$ is the same as $G^{\prime}$ given in Eq.~(\ref{eq:21}). It is clear that
the gain is proportional to the population inversion $J_{z}$ when
$\Delta=\omega_{d}-\omega_{c}=0$.
However, the gain becomes complicated when $\omega_{d}\neq\omega_{c}$. From Eq.~(\ref{eq:22}), we can obtain the threshold condition
$\gamma_{m}=G$, that is, the phonon can be amplified when $G
> \gamma_{m}$. We can see from Eq.~(\ref{eq:25-1}) that the threshold
condition in Ref.~\cite{Vahala2} is valid only when the cavity
field is resonantly driven.

\begin{figure}
\includegraphics[bb=0 190 540 665, width=4.2 cm, clip]{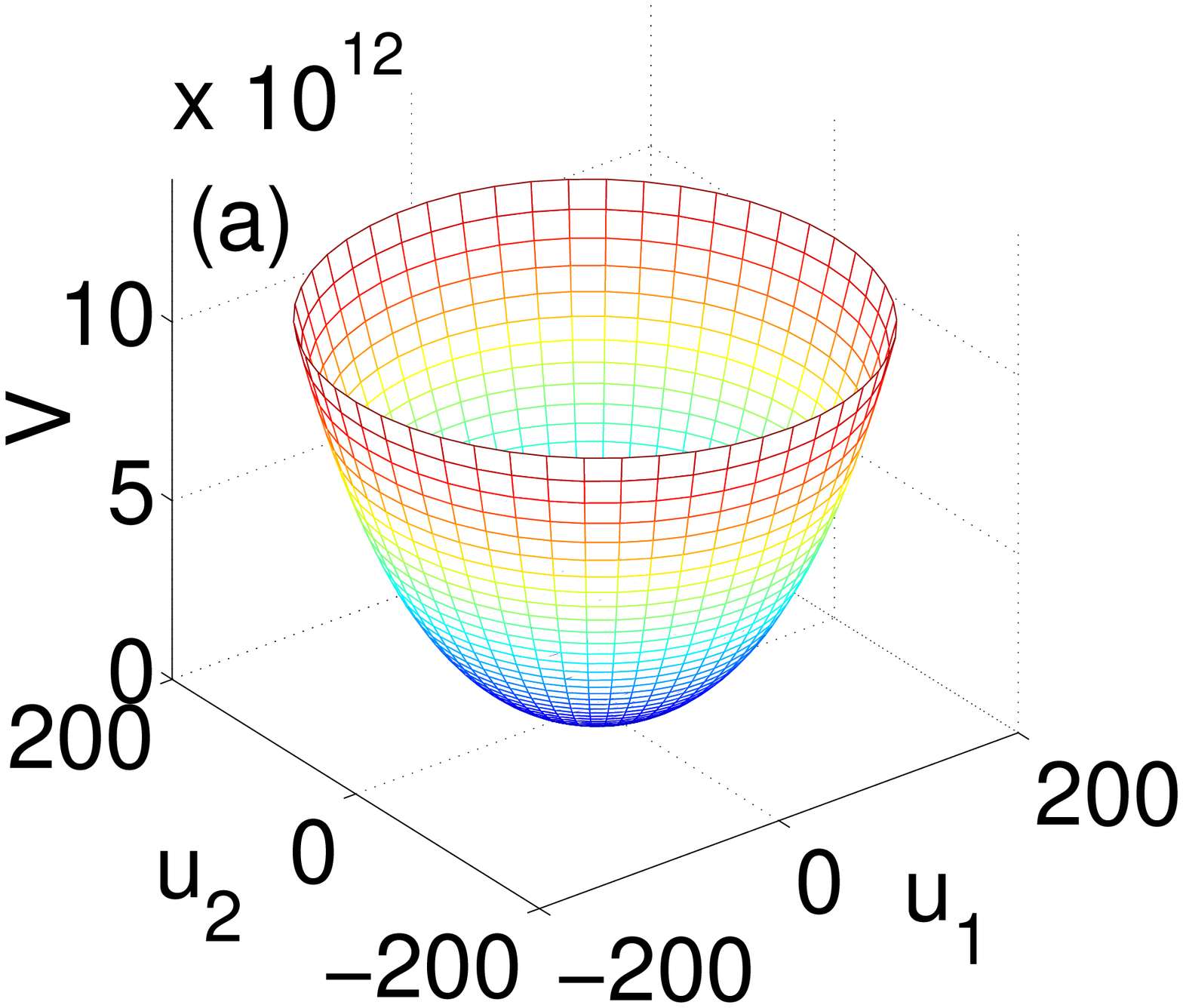}
\includegraphics[bb=0 190 540 665, width=4.2 cm, clip]{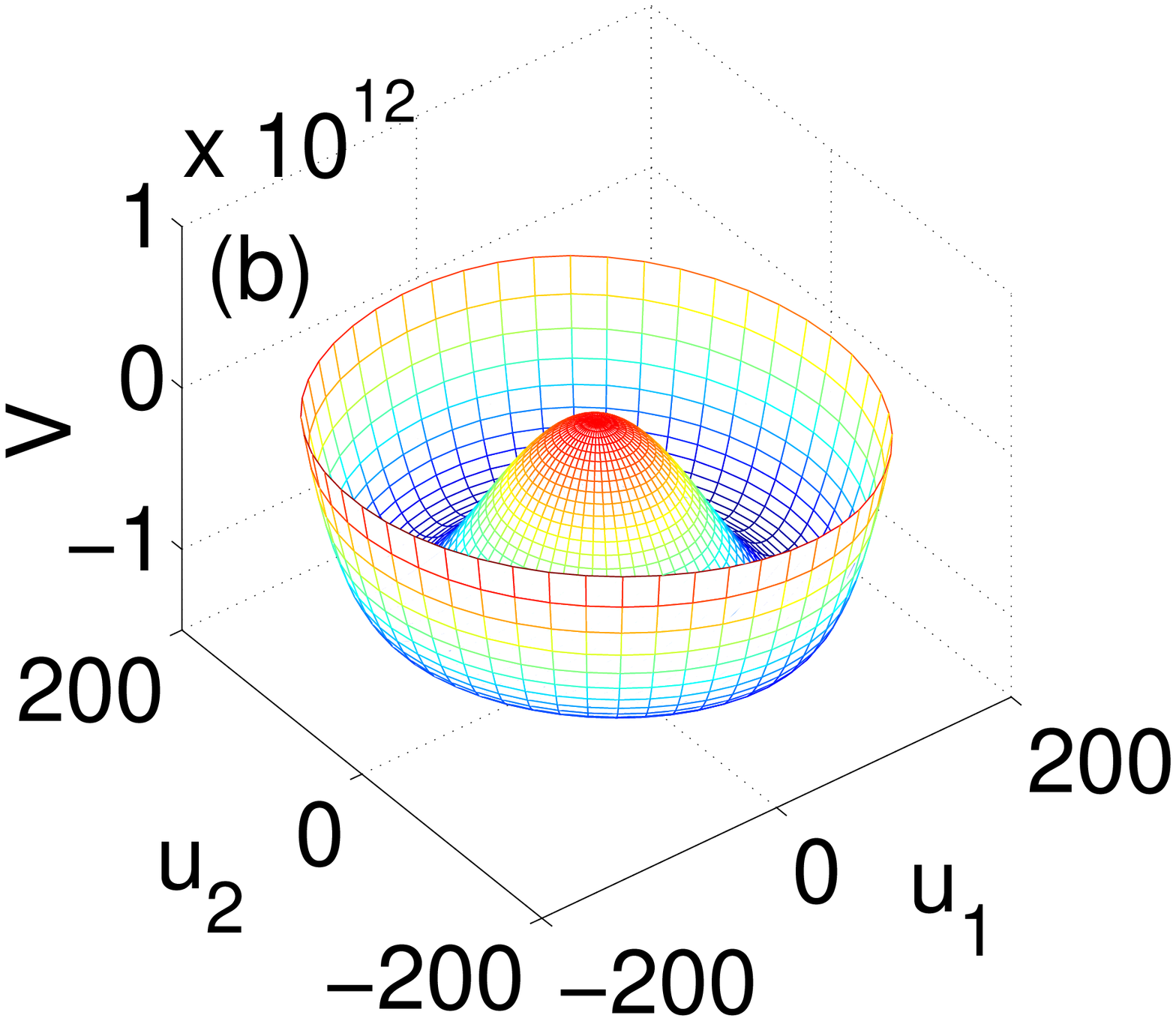}
\includegraphics[bb=0 190 540 665, width=4.2 cm, clip]{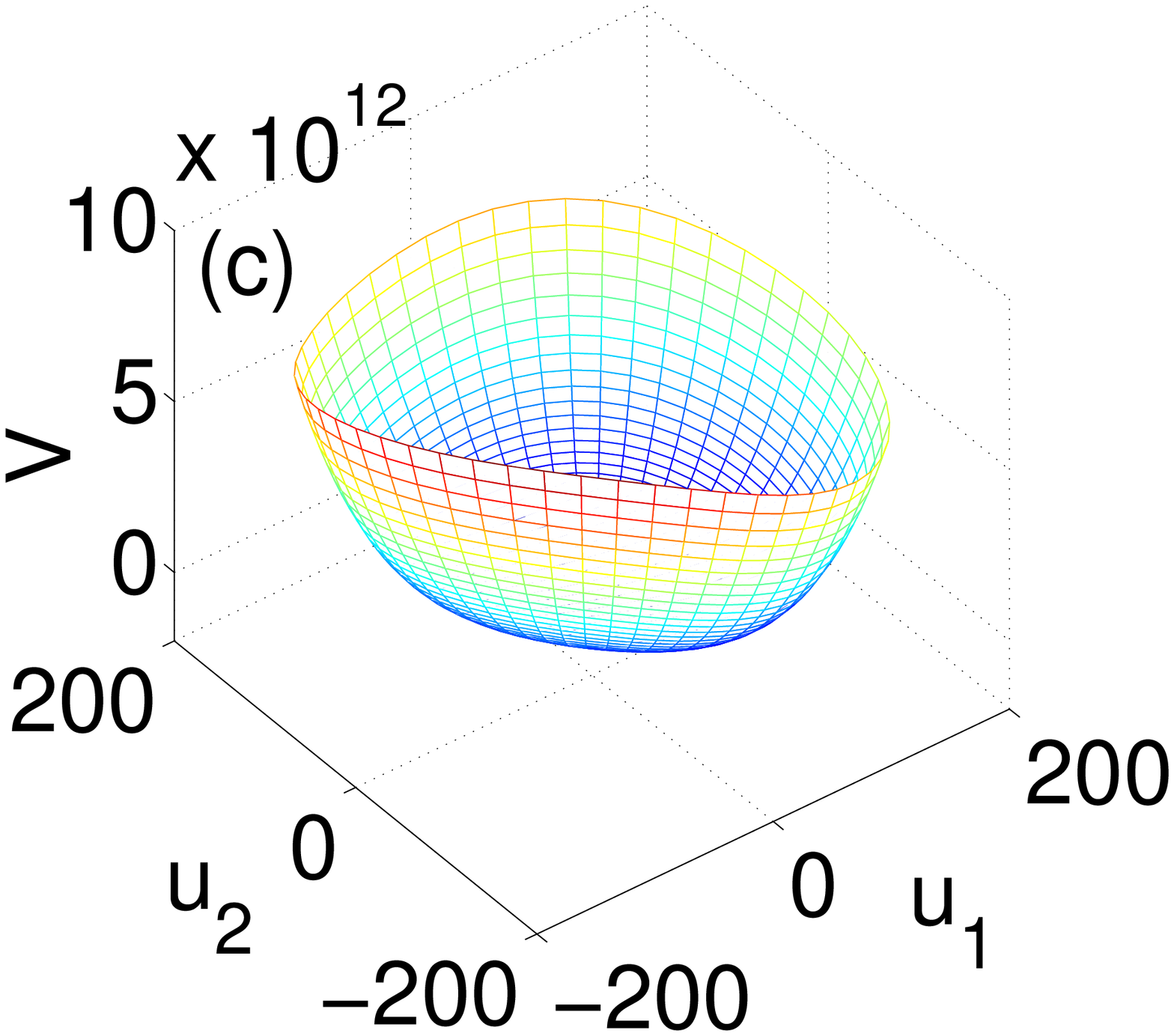}
\includegraphics[bb=0 190 540 665, width=4.2 cm, clip]{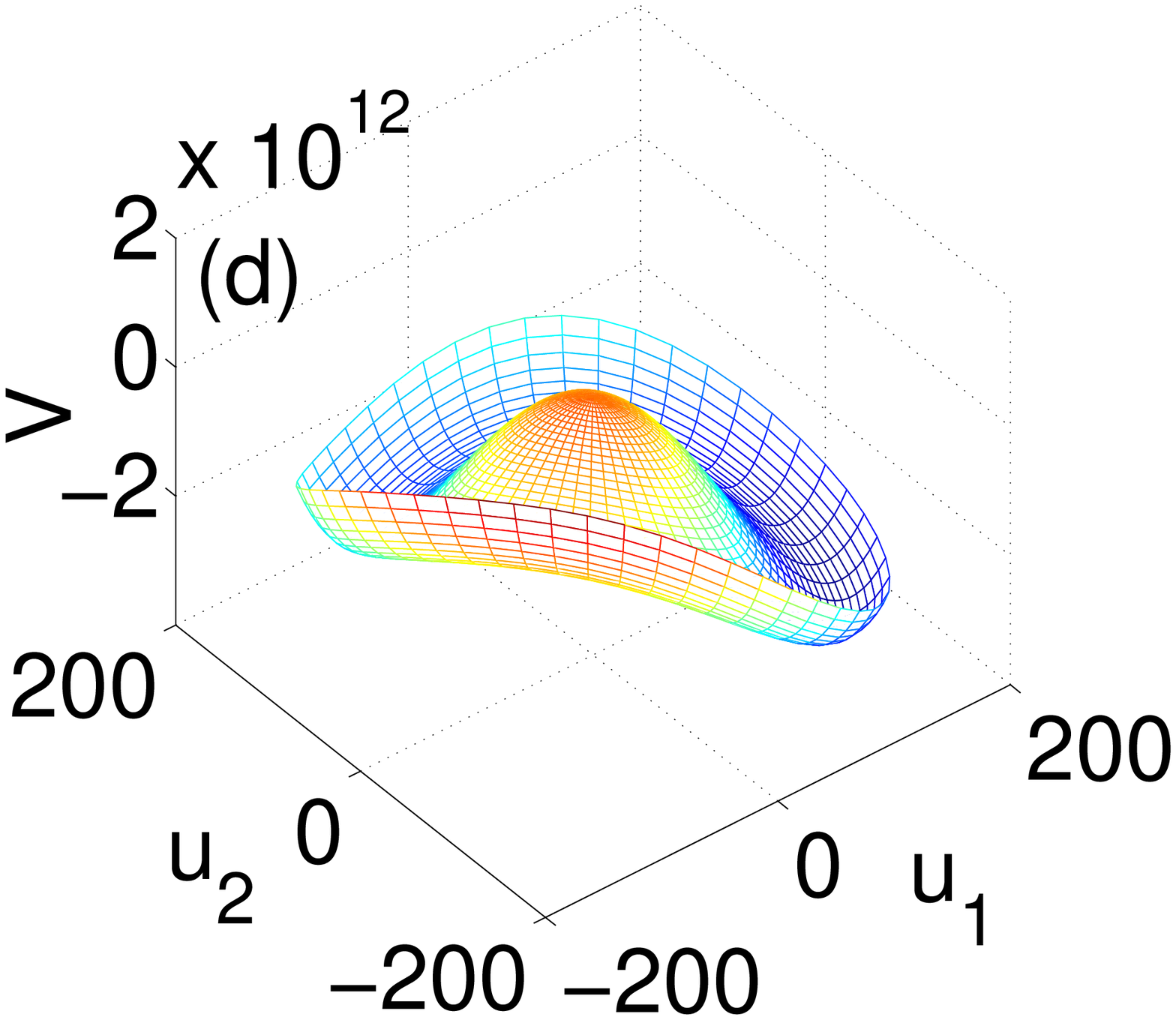}
\caption[]{(Color online) The two-dimensional scalar potential $V$
as functions of $u_{1}$ and $ u_{2} $. The parameters corresponding
to four figures are: (a) $\eta_{1}/(2\pi)=100$ MHz,
$\eta_{2}/(2\pi)=50$ MHz, $\alpha^{\prime}/(2\pi)=-50$ MHz,
$\text{Im}(G_{1})/(2\pi)=-1$ MHz , $\varepsilon_{6}/(2\pi)=2000$ Hz,
$\eta_{7}/(2\pi)=1800$ Hz, and $\eta_{8}/(2\pi)=10$ Hz; (b)
$\alpha^{\prime}/(2\pi)=36$ MHz, and the other parameters are the same as in (a);
(c) $\eta_{1}/(2\pi)=1$ GHz, $\eta_{2}/(2\pi)=0.5$ GHz,
$\alpha^{\prime}/(2\pi)=-5$ MHz, $\text{Im}(G_{1})/(2\pi)=-20$ MHz,
$\varepsilon_{6}/(2\pi)=0.02$ MHz, $\eta_{7}/(2\pi)=2500$ Hz, and
$\eta_{8}/(2\pi)=100$ Hz; (d) $\alpha^{\prime}/(2\pi)=20$ MHz, and the other
parameters are the same as in (c). }\label{fig4}
\end{figure}

\subsection{Phase diagram for phonon lasing}\label{3c}

To sketch the phase diagram of the phonon lasing, we further
express $J_{z}$ in Eq.~(\ref{eq:22}) as the variables of
mechanical mode by adiabatically eliminating the degrees of
freedom of the cavity mode. As a result we can express $J_{z}$ as the
function of phonon operators in the regime close to the threshold
such that $n\ll \left(\gamma^{2}_{c}+\delta^{2}\right)/4\chi^{2}$
(see Eq.~(\ref{eq:B3}) in Appendix~\ref{A2}).

By replacing $J_{z}$ in Eq.~(\ref{eq:20}) by
Eq.~(\ref{eq:B3}), we obtain
\begin{eqnarray}\label{eq:24}
\frac{d b}{d t}
&=&G_{0}+G_{1} b+G_{2}b^{\dagger}b+G_{3}b b-G_{4}b^{\dagger}bb,
\end{eqnarray}
up to third-order terms of the operator $b$ near the threshold
value of the phonon lasing. The coefficients $G_{i}$ (for $i=0,1\cdot\cdot\cdot,4$)
 can be found from Eqs.~(\ref{eq:B8}) to (\ref{eq:B12}) in Appendix~\ref{A2}.
By calculating the real part of $G_{1}$,  the net gain of phonon can be given as,
\begin{equation}\label{eq:25}
\alpha^{\prime}=\eta_{3}-\gamma_{m}.
\end{equation}
The definitions of $\eta_{i}$ (for $i=1,2\cdot\cdot\cdot,8$)
are defined in Eqs.~(\ref{eq:B13})-(\ref{eq:B20}) in Appendix~\ref{A2}.

In Fig.~\ref{fig3}, the variations of the gain $\alpha^{\prime}$ as a function
of the strength $|\Omega|$ of the driving field is plotted. If the value of
$\alpha^{\prime}$ is positive, the phonon lasing can be realized. Moreover,
 the phonon lasing also requires that the system is in the stable region.
From Fig.~\ref{fig3}(a), we find that the region with both $\alpha^{\prime}>0$
and ${\rm Re}(\lambda)<0$, which is for a stable phonon lasing, is smaller
for $\Delta=g$ than that for $\Delta=g/2$.  That is, the region between
$|\Omega_{1}|/(2\pi)$ and $|\Omega_{2}|/(2\pi)$ is smaller than that
between $|\Omega_{3}|/(2\pi)$ and $|\Omega_{4}|/(2\pi)$. From Fig.~\ref{fig3}(b),
we also find that the decay rate of the cavity field affects both the stability
and the gain. It is seen that smaller decay rate $\gamma_{c}$ of the cavity field
leads to a larger region for stable phonon lasing, that is,
the region between $|\Omega_{1}|/(2\pi)$ and $|\Omega_{4}|/(2\pi)$ corresponding
to $\gamma_{c}/(2\pi)=2.8$ MHz is larger than that between $|\Omega_{2}|/(2\pi)$
and $|\Omega_{3}|/(2\pi)$ corresponding to $\gamma_{c}/(2\pi)=4.8$ MHz.

With the semi-classical approximation, the
phonon field $b$ can be written as a two-dimensional vector
$b=(u_{1}\,\,\,u_{2})^{T}$ with $u_{1}=\text{Re}(b)$ and $u_{2}=\text{Im}(b)$.
Thus, we can obtain the dynamical equations for $u_{1}$ and
$u_{2}$ as shown in Eqs.~(\ref{eq:B21}) and (\ref{eq:B22}) of Appendix B. We should note here
that it is usually difficult to define a scalar potential for the mechanical mode.

We now discuss a special case with $\delta=0$. In this case, the equations of motion
for $u_{1}$ and $u_{2}$ are given as in Eqs.~(\ref{eq:B27}) and (\ref{eq:B28}).
When $u_{1}\gg u_{2}$, the terms containing $u_{2}$ are much
smaller than the same order terms including $u_{1}$, then we can
approximately define a two-dimensional scalar potential
 \begin{eqnarray}\label{eq:29}
V&\approx&-\eta_{1}u_{1}-\eta_{2}u_{2}+\text{Im}(G_{1})u_{1}u_{2}-\frac{\alpha^{\prime}}{2}
(u^{2}_{1}+u^{2}_{2})-\frac{\varepsilon_{6}}{3}u^{3}_{1}\nonumber\\
& &+\frac{\eta_{7}}{4}(u^{2}_{1}+u^{2}_{2})^{2}-\eta_{8}
u^{3}_{1}u_{2}-\frac{\eta_{8}}{3}u^{3}_{2}u_{1}.
\end{eqnarray}
The definition of $\varepsilon_{6}$ can be found in Eq.~(\ref{eq:B23}) in Appendix~\ref{A2}.
Actually, in other parameter regimes,
e.g., a larger $\delta$, we may also have $u_{2}\gg u_{1}$, and
the scalar potential similar to Eq.~(\ref{eq:29}) can also be
obtained.

In Fig~\ref{fig4}, the two-dimensional scalar potential $V$ is plotted as functions of
$u_{1}$ and $u_{2}$.  We find that when the coefficients
$\eta_{1}$, $\eta_{2}$ and $\varepsilon_{7}$ are negligibly small,
the potential $V$ approximately has a rotating symmetry in the
$u_{1}$-$u_{2}$ plane, otherwise, the symmetry is broken.
It is also easily found that the potential $V$ has only one
equilibrium point given at $b=(u_{1},u_{2})=0$ when
$\alpha^{\prime}< 0$, that is, there is no phonon lasing (In fact, it is possible to
have nonzero equilibrium points for $V$ if the value of $\alpha^{\prime}$
is negative but very close to $0$). However
if $\alpha^{\prime}> 0$, $b=0$ is not a stable point.
Instead, two new stable points appear. We find that these two
stable points are not symmetric for the nonzero  coefficients
of odd terms of $u_{1}$ and $u_{2}$  (in Eq.~(\ref{eq:29}))
  in contrast to the case of photon lasing~\cite{Haken}
in which two stable points are always symmetric. We note that
Eq.~(\ref{eq:24}) is valid only near the threshold.

\begin{figure}
\includegraphics[bb=60 180 490 600, width=8 cm, clip]{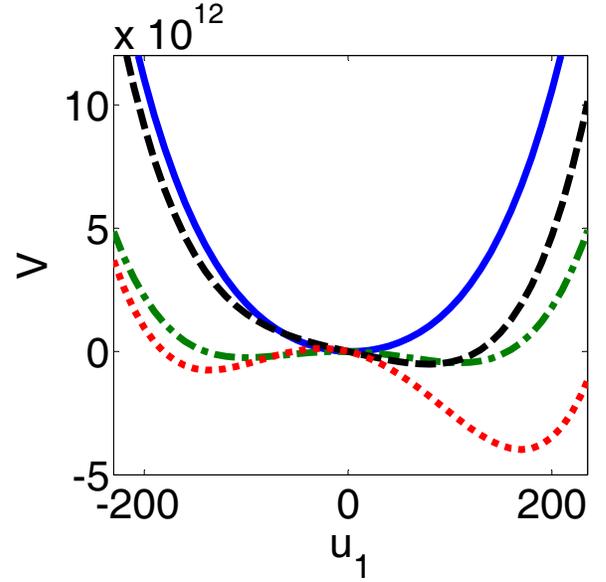}
\caption[]{(Color online) One dimensional potential $V$ as a
function of $u_{1}$. The parameters corresponding to four curves
are: (a) $\eta_{1}/(2\pi)=100$ MHz, $\alpha^{\prime}/(2\pi)=-50$ MHz ,
$\varepsilon_{6}/(2\pi)=2000$ Hz, and $\eta_{7}/(2\pi)=1800$ Hz (blue solid curve); (b)
$\alpha^{\prime}/(2\pi)=36$ MHz, the other parameters are the same as in (a) (green dash-dotted curve);
(c) $\eta_{1}/(2\pi)=1$ GHz, $\alpha^{\prime}/(2\pi)=-5$ MHz,
$\varepsilon_{6}/(2\pi)=0.02$ MHz, and $\eta_{7}/(2\pi)=2500$ Hz  (black dashed curve); (d)
$\alpha^{\prime}/(2\pi)=20$ MHz, the other parameters are the same as in
(c) (red dotted curve). }\label{fig5}
\end{figure}

In the case that the terms including $u_{2}$ in $V$ are negligibly smaller than
the terms including $u_{1}$, then we can neglect all terms
including $u_{2}$ ($u_{1} \gg u_{2}$) in Eq.~(\ref{eq:29}) and
approximately define the one-dimensional scalar potential
\begin{eqnarray}\label{eq:28}
V=-\eta_{1}u_{1}-\frac{\alpha^{\prime}}{2}u^{2}_{1}-\frac{\varepsilon_{6}}{3}u^{3}_{1}+\frac{\eta_{7}}{4}u^{4}_{1}.
\end{eqnarray}
In Fig.~\ref{fig5}, $V$ is plotted as a
function of $u_{1}$ according to Eq.~(\ref{eq:28}). We also show in
Fig.~\ref{fig5} that the shape of the potential function $V$
changes from the parabolic potential well to two symmetric
potential wells, and then to two asymmetric potential wells as the parameters are varied.
 Actually, potential similar to that given in Fig.~\ref{fig5} can also be found
for $u_{2}$ in some parameter regimes.

\section{Statistical properties of the phonon}

We now study the statistical properties of the phonons by
calculating the second-order degrees of coherence in the same way
as usually done for photons~\cite{scully}. The normalized
equal-time second-order correlation function of the phonons is defined
as $g_{2}(0)=\langle b^{\dag}(t)b^{\dag}(t)b(t)b(t)\rangle/\langle
b^{\dag}(t)b(t)\rangle^{2}$. Using the small-fluctuation
approximation as shown in Eqs.~(\ref{eq:10})-(\ref{eq:12}), the degree of
second-order coherence can be written as
\begin{eqnarray}\label{eq:30a}
g_{2}(0)&=&\frac{|B_{0}|^{4}+2\text{Re}\left[B^{\ast2}_{0} \langle\beta(t)\beta(t) \rangle\right]+4|B_{0}|^{2} \langle\beta^{\dagger}(t)\beta(t) \rangle}
{\left(|B_{0}|^{2}+ \langle\beta^{\dagger}(t)\beta(t) \rangle\right)^{2}}\nonumber\\
& +&\frac{ \langle\beta^{\dagger}(t)\beta^{\dagger}(t)\beta(t)\beta(t) \rangle}{\left(|B_{0}|^{2}+ \langle\beta^{\dagger}(t)\beta(t) \rangle\right)^{2}}.
\end{eqnarray}
If the phonon mode is in the coherent state, the value of $g_{2}(0)=1$.

It is somewhat difficult to calculate the fluctuation operator
$\beta(t)$ in the time domain, thus we try to calculate it in the
frequency domain. By introducing the Fourier transform, we can obtain the
dynamical equation for the fluctuation operators in the frequency
domain as in Eqs.~(\ref{eq:C4})-(\ref{eq:C6}) of Appendix~\ref{A3}.
In this case, we can obtain the phonon fluctuation operators in the frequency domain
as follows
\begin{eqnarray}\label{eq:59}
\tilde{\beta}(\omega)&=&p_{1}(\omega) \tilde{b}_{\rm in}(\omega)
+p_{2}(\omega) \tilde{b}^{\dagger}_{\rm in}(\omega)+p_{3}(\omega)\tilde{\Gamma}_{1}(\omega)\nonumber\\
&+ &p_{4}(\omega)\tilde{\Gamma}^{\dagger}_{1}(\omega)
+p_{5}(\omega)\tilde{\Gamma}_{2}(\omega)
+p_{6}(\omega)\tilde{\Gamma}^{\dagger}_{2}(\omega).
\end{eqnarray}
The parameters $p_{i}$ ($i=1, 2,\cdot\cdot\cdot, 6$) in
Eq.~(\ref{eq:59}) are given from Eqs.~(\ref{eq:C7})-(\ref{eq:C12}) in Appendix~\ref{A3}. From
Eqs.~(\ref{eq:n1})-(\ref{eq:n4}), the correlation functions
of the input noise operators in the frequency domain can be easily obtained.
Then, from Eq.~(\ref{eq:59}), we can obtain the
correlation functions of the phonon fluctuation operators in the
frequency domain as follows
\begin{eqnarray}
\langle\tilde{\beta}(\omega)\tilde{\beta}(\omega^{\prime})\rangle&=&
2\pi\Gamma_{\rm{\beta \beta}}\delta(\omega+\omega^{\prime}),\\
\langle\tilde{\beta}^{\dagger}(\omega)\tilde{\beta}(\omega^{\prime})\rangle&=&
2\pi \Gamma_{\rm{\beta^{\dagger}\beta}}\delta(\omega+\omega^{\prime}).
\end{eqnarray}
with the coefficients defined as
\begin{eqnarray}
\Gamma_{\rm {\beta \beta}}(\omega)&=&p_{1}(\omega)p_{2}(-\omega)(n_{b}+1) +p_{2}(\omega)p_{1}(-\omega)n_{b}\nonumber\\
&+ &p_{3}(\omega)p_{4}(-\omega) \gamma_c +p_{5}(\omega)p_{6}(-\omega)\gamma_c,\\
\Gamma_{\rm {\beta^{\dagger}\beta}}(\omega)&=&|p_{1}(-\omega)|^{2}n_{b}+|p_{2}(-\omega)|^{2}(n_{b}+1)\nonumber\\
& +& |p_{4}(-\omega)|^{2}\gamma_c+|p_{6}(-\omega)|^{2}\gamma_c.
\end{eqnarray}
Here, $\Gamma_{\rm {\beta \beta}}(\omega)$ and
$\Gamma_{\rm {\beta^{\dagger}\beta}}(\omega)$ represent the correlation
spectra of the phonon fluctuation operators. If we assume that the environmental
noises have Gaussian distributions, then, from Wick's theorem, all
higher-order correlation functions can be written as the
combinations of the first- and second-order correlation
functions~\cite{girvin}. Thus, with straightforward but tedious
calculations, we have
$\langle\beta^{\dagger}(t)\beta^{\dagger}(t)\beta(t)\beta(t)
\rangle=2Y^{2}_{\rm{\beta^{\dagger}\beta}}+
|Y_{\rm{\beta\beta}}|^{2}$, where $Y_{\rm{\beta\beta}}$ and
$Y_{\rm{\beta^{\dagger}\beta}}$ can be calculated as
\begin{eqnarray}
Y_{\rm{\beta\beta}}&=&\langle\beta(t)\beta(t)\rangle=\frac{1}{2\pi}\int^{+\infty}_{-\infty} \Gamma_{\beta\beta}(\omega)d\omega,\\
Y_{\rm{\beta^{\dagger}\beta}}&=&\langle\beta^{\dagger}(t)\beta(t)\rangle=\frac{1}{2\pi}\int^{+\infty}_{-\infty} \Gamma_{\beta^{\dagger}\beta}(\omega)d\omega.
\end{eqnarray}

\begin{figure}
\includegraphics[bb=65 220 460 615, width=4.25 cm, clip]{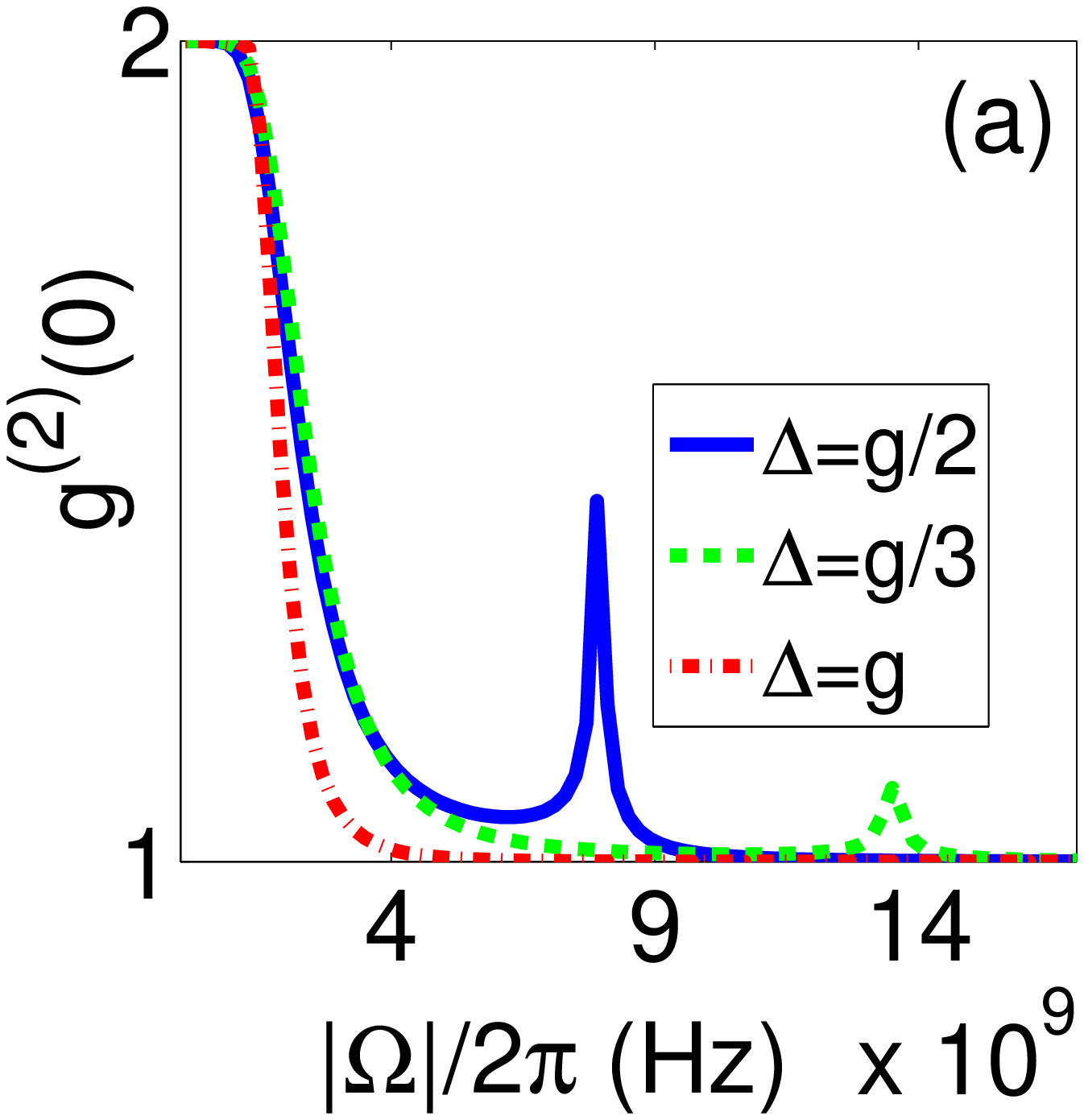}
\includegraphics[bb=45 220 440 615, width=4.25 cm, clip]{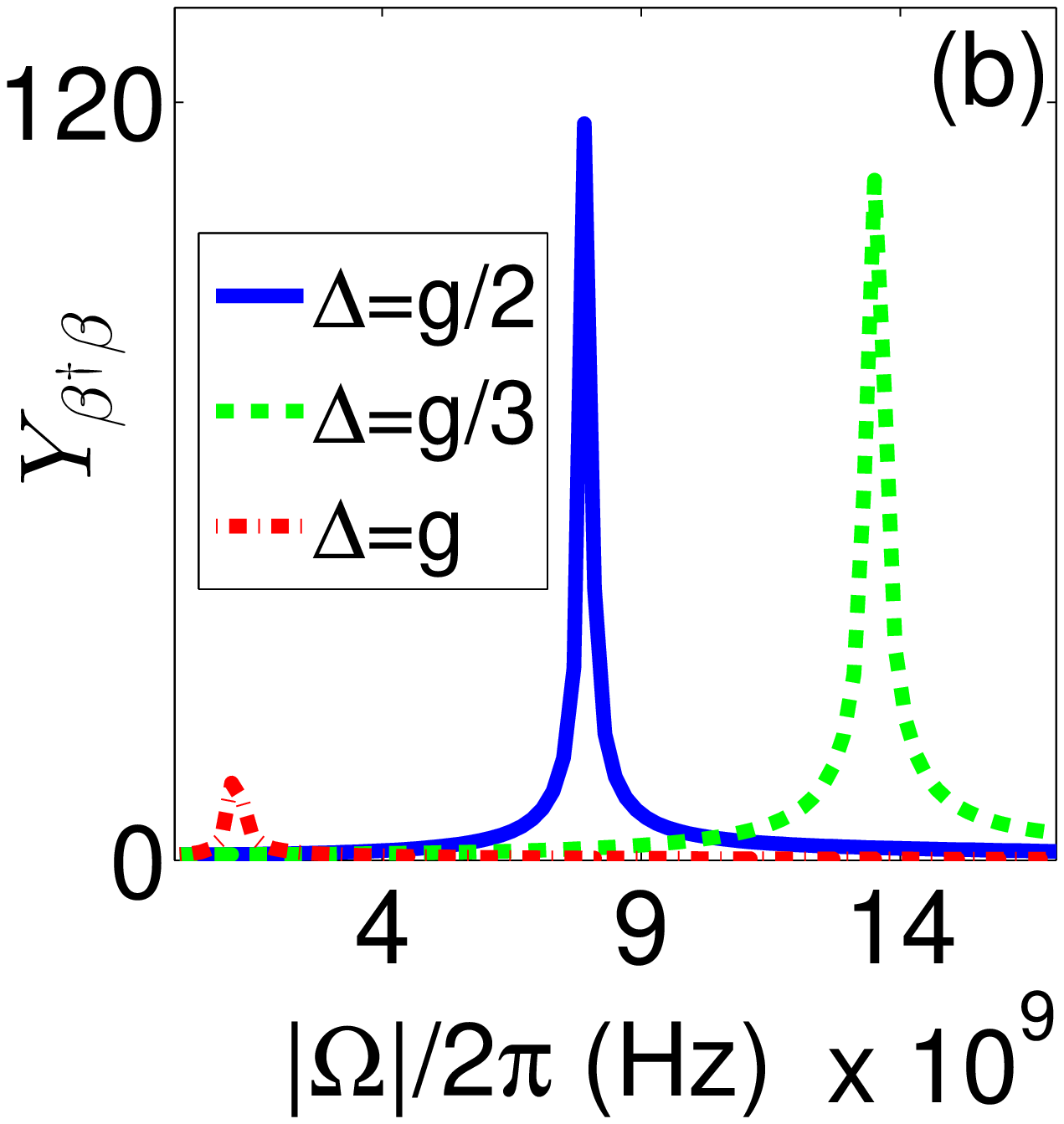}
\caption[]{(Color online) The degree of second-order coherence $g^{(2)}(0)$ in (a) and
the phonon number fluctuation $Y_{\rm{\beta^{\dagger}\beta}}$ in (b) as the
functions of the strength $ |\Omega| $ of the driving field.
The three curves correspond to  different detunings:
(1) $\Delta=g/2$ (blue solid curve);
(2) $\Delta=g/3$ (green dashed curve);
(3) $\Delta=g$ (red dash-dotted curve).
 The other parameters are the same as in Fig.~\ref{fig2} except $T=1$ mK.}\label{fig6}
\end{figure}

\begin{figure}
\includegraphics[bb=30 210 455 625, width=4.27 cm, clip]{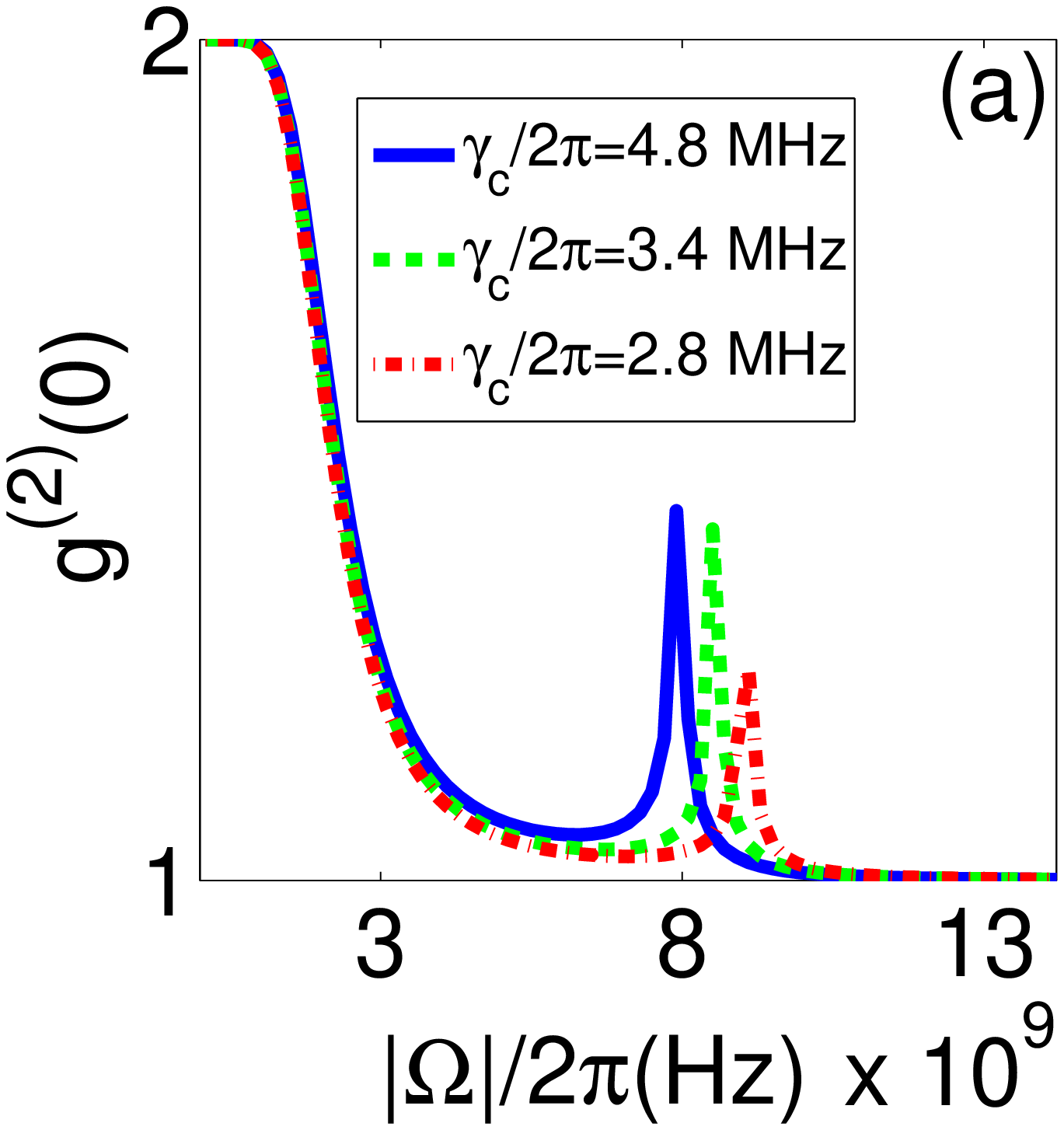}
\includegraphics[bb=25 210 455 630, width=4.27 cm, clip]{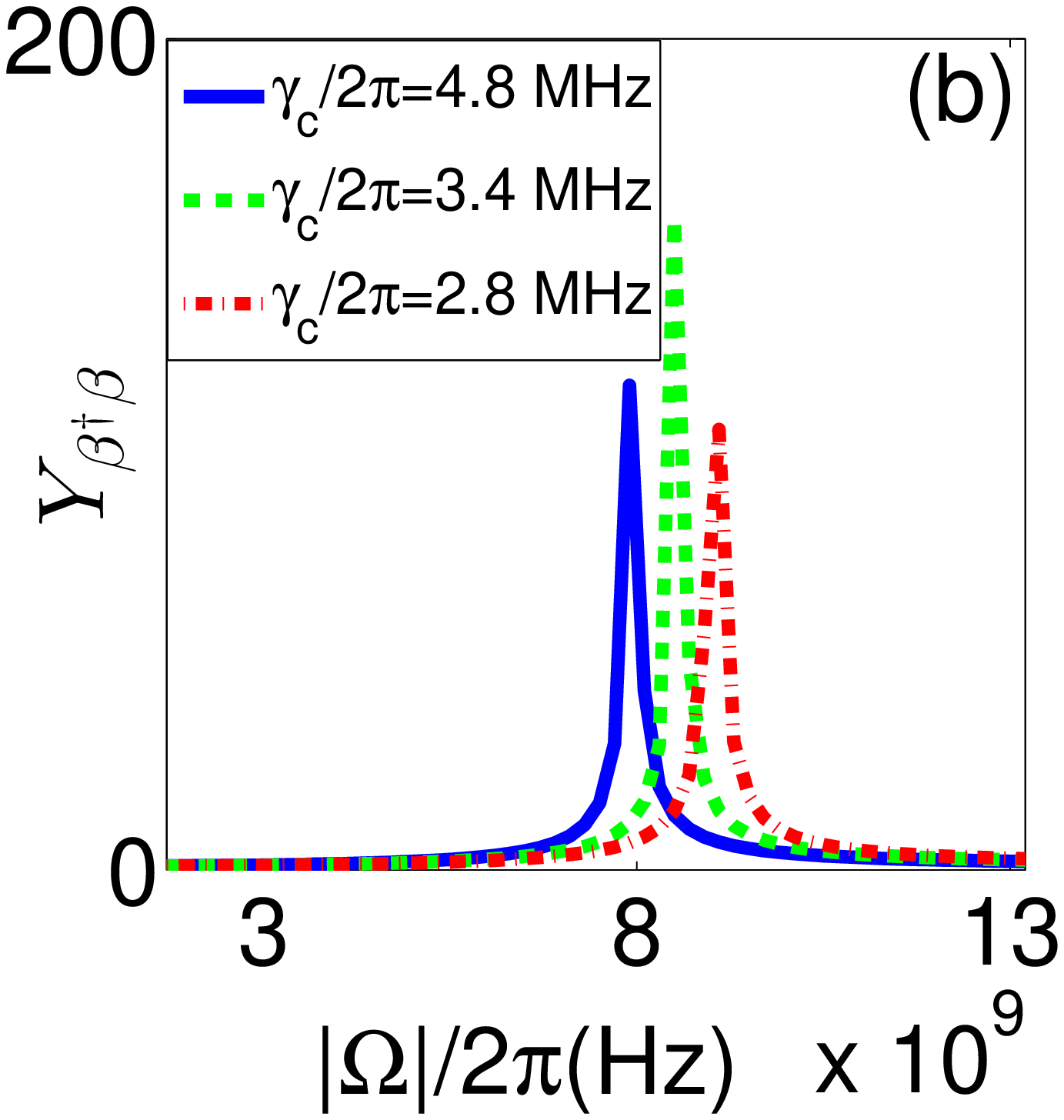}
\caption[]{(Color online) The degree of second-order coherence $g^{(2)}(0)$ in (a) and
the phonon number fluctuation $Y_{\rm{\beta^{\dagger}\beta}}$ in (b) as the
function of the strength $ |\Omega| $ for the driving field.
The three curves correspond to  different cavity decay rates:
(1) $\gamma_{c}/(2\pi)=4.8$ MHz (blue solid curve);
(2) $\gamma_{c}/(2\pi)=3.4$ MHz (green dashed curve);
(3) $\gamma_{c}/(2\pi)=2.8$ MHz (red dash-dotted curve).
The other parameters are the same as in Fig.~\ref{fig2}
 except $\Delta=g/2$ and $T=1$ mK.}\label{fig7}
\end{figure}

\begin{figure}
\includegraphics[bb=40 210 475 630, width=4.25 cm, clip]{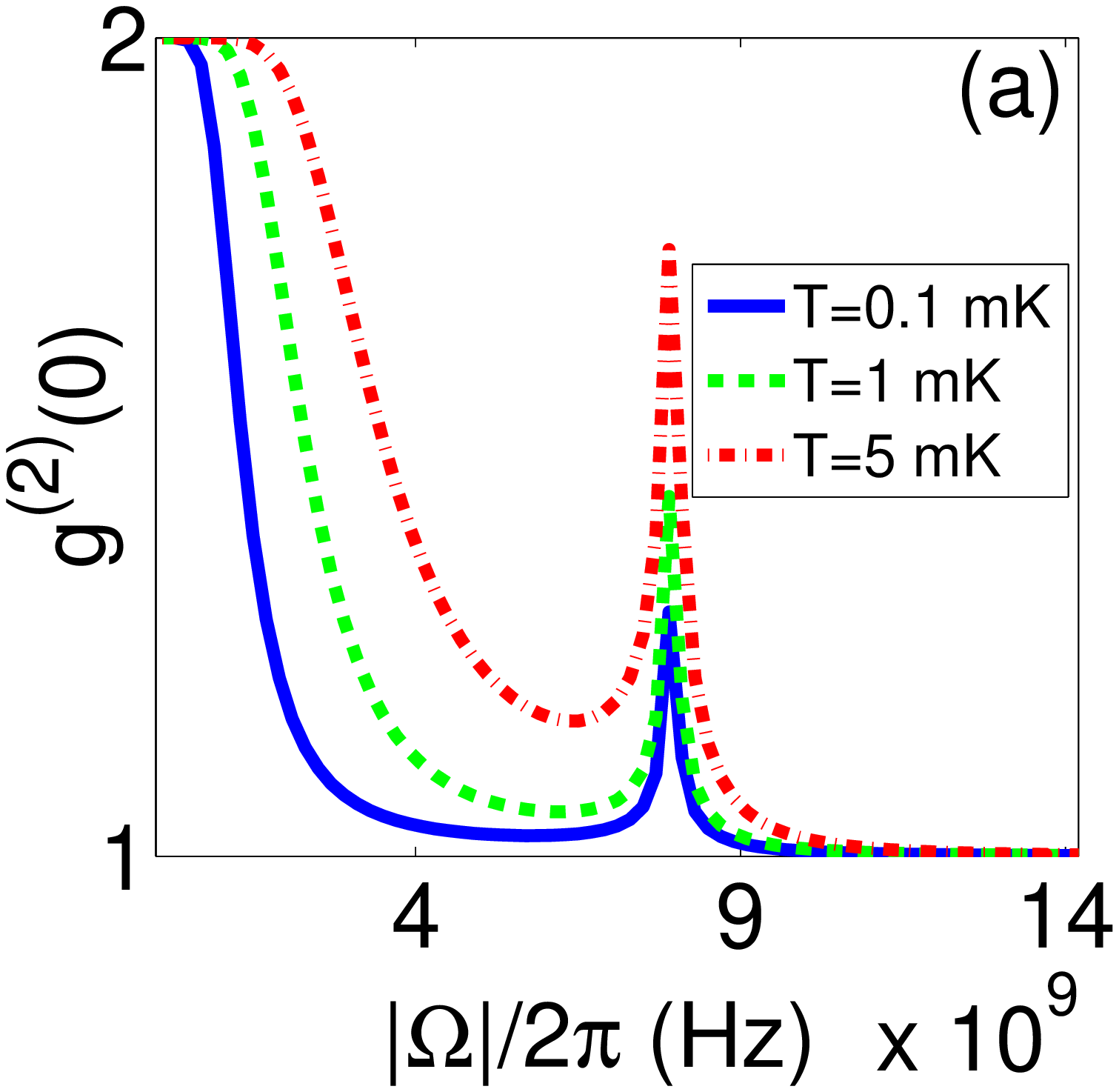}
\includegraphics[bb=90 205 505 615, width=4.25 cm, clip]{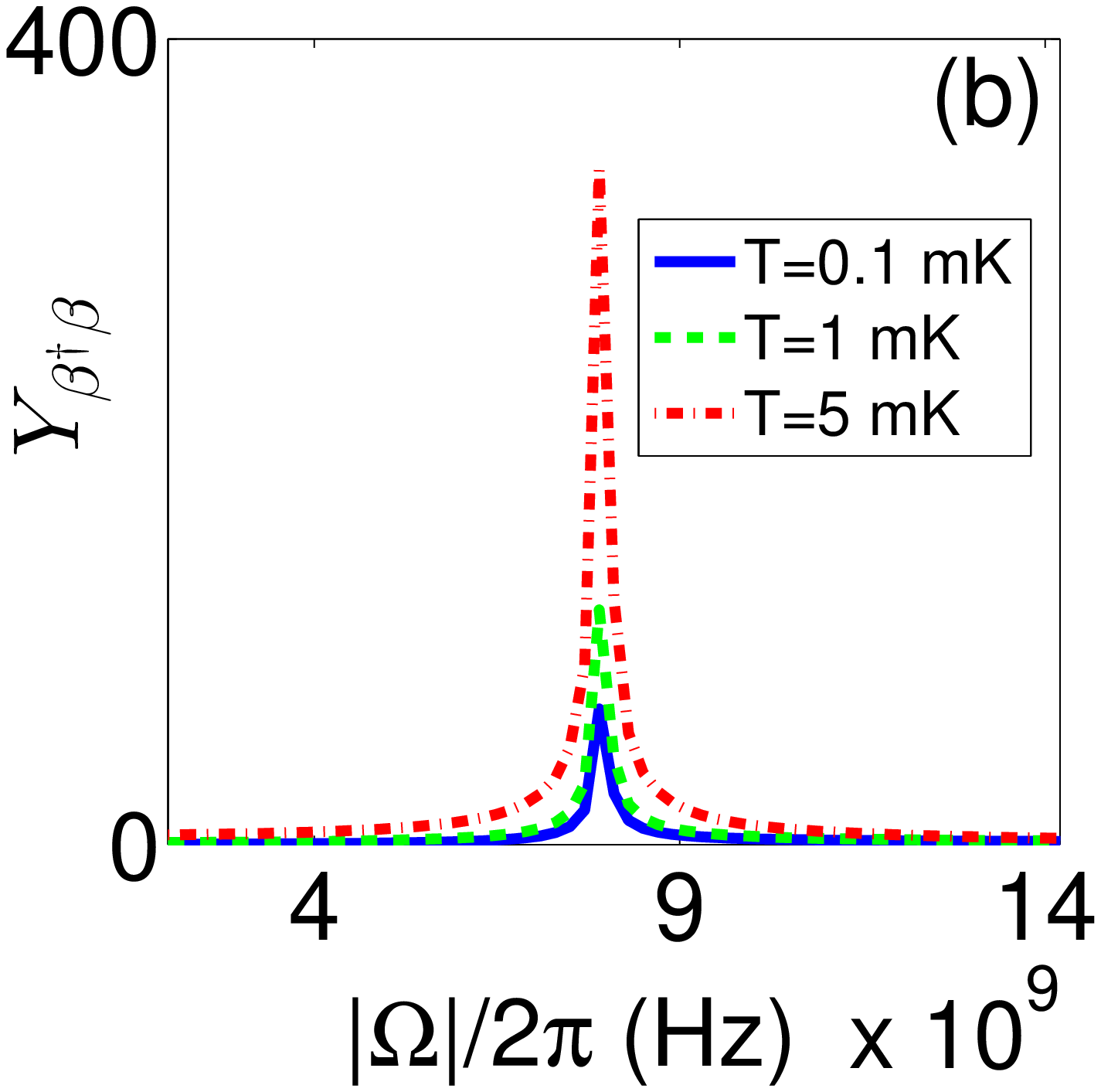}
\caption[]{(Color online) The degree of second-order coherence $g^{(2)}(0)$ in (a) and
the phonon number fluctuation $Y_{\rm{\beta^{\dagger}\beta}}$ in (b) as the
function of the strength $ |\Omega| $ for the driving field.
 The three curves correspond to  different temperatures:
 (1) $T=0.1$ mK (blue solid curve);
 (2) $T=1$ mK (green dashed curve);
 (3) $T=5$ mK (red dash-dotted curve).
 The other parameters are the same as in Fig.~\ref{fig2}
 except $\Delta=g/2$.}\label{fig8}
\end{figure}

The degree of second-order coherence $g^{(2)}(0)$  for the phonons is plotted as a function
of the strength $\Omega$ of the driving field for different detunings $\Delta$ ,  cavity decay rates $\gamma_{c}$,
and temperatures $T$ in Fig.~\ref{fig6}(a), Fig.~\ref{fig7}(a), and Fig.~\ref{fig8}(a), respectively.
 All three figures show $g^{(2)}(0)=2$ when the driving
field is not applied to the cavity. This meas that the phonons are in the thermal equilibrium state. However,
when the driving field strength $\Omega$ approaches to (or is above) the threshold, the degree of coherent $g^{(2)}(0)$
quickly approaches to (or is) $1$, which means that the phonon is in a coherent state and the phonon lasing occurs.

In contrast to $g^{(2)}(0)=1$ in the photon lasing with two-level atoms~\cite{Haken}, we find that there
is a resonant peak in the curve of $g^{(2)}(0)$, the reason is that the spectrum $\Gamma_{\rm
{\beta^{\dagger}\beta}}(\omega)$ of phonon fluctuation operators is not only the
function of frequency $\omega$, but also the functions of the strength $\Omega$ for the driving
field, cavity decay rate $\gamma_{c}$, the detuning $\Delta$ and so on.
When the strength $\Omega$ of the driving field varies, the resonant peaks will appear in
the spectrum $\Gamma_{\rm {\beta^{\dagger}\beta}}(\omega)$, and then in the degree of second-order coherence.
The correlation function $Y_{\rm{\beta^{\dagger}\beta}}$, which is the inverse Fourier transform of
$\Gamma_{\rm {\beta^{\dagger}\beta}}(\omega)$, is plotted as a function of strength $\Omega$
of the driving field for different detunings $\Delta$ in Fig.~\ref{fig6}(b),
cavity decay rates $\gamma_{c}$ in Fig.~\ref{fig7}(b)
and temperatures $T$ in Fig.~\ref{fig8}(b).

From Fig.~\ref{fig6} and Fig.~\ref{fig7},  we find that the detuning $\Delta$, cavity decay $\gamma_{c}$,
and temperature $T$ affect both the degree of second-order coherence and the correlation spectrum
of phonon fluctuation operators. We also find that the positions and heights of the resonant peaks of
$g^{(2)}(0)$ and $Y_{\rm{\beta^{\dagger}\beta}}$  change
when the parameters $\gamma_{c}$ and $\Delta$ are varied. The
temperature $T$, on the other hand, only affects the height of resonant peaks as shown in
Fig.~\ref{fig8}, and has no effect on their positions.
With the increase of the temperature, the mechanical mode is gradually thermalized,
and the coherent properties of
phonon state degrade.

\section{Conclusions}

In summary, we have theoretically analyzed the phonon
lasing studied in Ref.~\cite{Vahala2} for a coupled optomechanical system.
We have showed that in the steady-state the phonon mode shows unconventional
bistability in the regime of strong driving field.
We have derived the equations of phonon lasing by adiabatically
eliminating the cavity modes, and then obtained the gain and threshold of the mechanical amplification.
We have found that the positive net gain of phonon lasing can be obtained in a stable regime,
and thus it is possible to create stable coherent phonons in a coupled optomechanical system.
In particular, we have clarified that the threshold given in Ref.~\cite{Vahala2}
 is just a special case for the phonon lasing in coupled optomechanical systems.

Interestingly, our study shows that the phonon lasing generated by photonic supermodes is somewhat different from the photon lasing
generated by a two-level atomic system~\cite{scully}. For example, even
when there is no photon population inversion, the mechanical mode can
be amplified. In contrast to the photon lasing, there is no saturation for phonon lasing.
From the phase diagram of phonon lasing, we have found that a scalar potential can be approximately defined in some
special cases. Different from photon lasing~\cite{scully}, the symmetry of the effective
potential for phonon mode can be broken when the driving field is strong enough.

We have also studied the degree of second-order coherence of the
phonon mode. We have showed that it tends to one when the driving field is strong enough.
This means that the phonon mode reaches to a coherent
state. Moreover, a resonant peak occurs in the degree of second-order coherence when
the strength of the driving field satisfies certain conditions. This distinguished difference from the
photon lasing~\cite{scully} is due to the dependence of phonon fluctuation operators on the properties of the driving field.
Our study have clarified some important points for phonon lasing in coupled optomechanical systems.
We believe that our results will be useful in designing new experiments and in interpreting their results.
 In particular, they will be of great interest for the efforts to demonstrate phonon lasing
 in parity-time (PT-) symmetric optical and  optomechanical systems~\cite{Peng,Jing,Li}.

\section{Acknowledgement}
This work is partially supported by  the National Natural Science Foundation
of China under Grant No. 61328502. Y.X.L. is supported by the NSFC under Grant
No. 61025022 and the National Basic Research Program of China Grant No. 2014CB921401. J.Z. is supported by
the NFSC under Grant Nos. 61174084, 61134008, 60904034.  L.Y. and S.K.O. are partially
 supported by ARO grant No. W911NF-12-1-0026.

\appendix

\section{Derivation of the dynamical equations}\label{A1}

We consider a multi-mode optomechanical system consisting of two
optical modes and one mechanical mode. The total Hamiltonian
including the environmental noises can be written as follows
\begin{eqnarray}
H&=&\hbar\omega_{a}a^{\dagger}_{L}a_{L}+\hbar\omega_{a}a^{\dagger}_{R}a_{R}+\hbar g\left(a^{\dagger}_{L}a_{R}+a^{\dagger}_{R}a_{L}\right)\nonumber\\
& &+\hbar\omega_{m} b^{\dagger}b-\hbar\chi\left(a^{\dagger}_{L}a_{L}-a^{\dagger}_{R}a_{R}\right)(b^{\dagger}+b)\nonumber\\
& &+i\hbar\left[\Omega \exp{(-i\omega_{d}t)}a^{\dagger}_{L}-h.c.\right]\nonumber\\
& &+\hbar\sum_{i}{\omega_{\rm bi}b^{\dagger}_{i}b_{i}}+\hbar\sum_{i}{\left(g_{\rm bi}bb^{\dagger}_{i}+h.c.\right)}
\nonumber\\
& &+\hbar\sum_{i}{\omega^{(L)}_{i}a^{\dagger}_{\rm Li}a_{\rm Li}}+\hbar\sum_{i}{\left(g^{(L)}_{i}a^{\dagger}_{\rm Li}a_{L}
+h.c.\right)}\nonumber\\
& &+\hbar\sum_{i}{\omega^{(R)}_{i}a^{\dagger}_{\rm Ri}a_{\rm Ri}}+\hbar\sum_{i}{\left(g^{(R)}_{i}a^{\dagger}_{\rm Ri}a_{R}+h.c.\right)}\qquad
\end{eqnarray}
$b^{\dagger}$ (or $b$) is the creation (or annihilation) operator
of the phonon mode with frequency $\omega_{m} $. $a^{\dagger}_{L}$
(or $ a_{L} $) and $a^{\dagger}_{R} $ (or $a_{R}$) represent the
creation (or annihilation) operators of the left and right cavity
fields respectively, and the corresponding frequencies of the two
bare cavities are both $\omega_{a}$. $ \chi $ is the coupling
strength between the cavity field and the mechanical resonator,
and the interaction strength between the two optical cavities is
$g$. $b^{\dagger}_{i}$ (or $b_{i}$) is the creation (or
annihilation) operator of the $i$-th environmental noise mode
coupled with the mechanical resonator with coupling strength
$g_{\rm bi}$. $a^{\dagger}_{Li}$ (or $a_{\rm Li}$) and
$a^{\dagger}_{Ri}$ (or $a_{\rm Ri}$) are the creation (or
annihilation) operators of $i$-th environmental noise mode coupled
with the optical modes in the left and right cavities with
coupling strengthes $g^{(L)}_{i}$ and $g^{(R)}_{i}$. A classical
driving field with frequency $\omega_{d}$ and amplitude $\Omega $
is injected into the left cavity. $ \Delta=\omega_d-\omega_a$ is
the detuning frequency between the diving field and the mode in
the left cavity (bare frequency).

We redefine two new optical modes $a_{1}=(a_{L}+a_{R})/\sqrt{2} $
and $a_{2}=(a_{L}-a_{R})/\sqrt{2} $, then in a rotating reference frame
given by the unitary operation
\begin{equation}
U=\exp\left\{-i\omega_{d}\left[\sum_{i=1}^{2}a^{\dagger}_{i}a_{i}
+\sum_{i}\left({a^{\dagger}_{Li}a_{Li}}+{a^{\dagger}_{Ri}a_{Ri}}\right)\right]t\right\},
\end{equation}
the Hamiltonian can be rewritten as
\begin{eqnarray}
H_{r}&=& H_{0}+\hbar\sum_{i}{\omega_{\rm bi}b^{\dagger}_{i}b_{i}}+\hbar\sum_{i}{\left(g_{\rm bi}bb^{\dagger}_{i}+h.c.\right)}\nonumber\\
& &+\hbar\sum_{i}{\omega^{(L)}_{i}a^{\dagger}_{\rm Li}a_{\rm Li}}+\hbar\sum_{i}{\omega^{(R)}_{i}a^{\dagger}_{\rm Ri}a_{\rm Ri}}\nonumber\\
& &+(\hbar/\sqrt{2})\sum_{i}{\left[g^{(L)}_{i}a^{\dagger}_{\rm Li}(a_{1}+a_{2})
+h.c.\right]}\nonumber\\
& &+(\hbar/\sqrt{2})\sum_{i}{\left[g^{(R)}_{i}a^{\dagger}_{\rm Ri}(a_{1}-a_{2})+h.c.\right]},\label{eq:A2}
\end{eqnarray}
with the Hamiltonian $H_{0}$ given in Eq.~(\ref{eq:1}). Here we also use the rotating wave approximation.

With the Hamiltonian in Eq.~(\ref{eq:A2}), we can obtain the
Heisenberg-Langevin equations of the total system as follows
\begin{eqnarray}
\dot{ a_{1}}&=&-i\left(g-\Delta\right)a_{1}+i\chi a_{2}b+\Omega /\sqrt{2}\nonumber\\ \label{eq:A3}
& &-(i/\sqrt{2})\sum_{i}{\left[g^{(L)\ast}_{i}a_{\rm Li}+g^{(R)\ast}_{i}a_{\rm Ri}\right]},\qquad\\ \label{eq:A4}
\dot{a_{2}}&=&i\left(g+\Delta\right)a_{2}+i\chi a_{1}b^{\dagger}+\Omega/\sqrt{2}\nonumber\\ \label{eq:A5}
& &-(i/\sqrt{2})\sum_{i}{\left[g^{(L)\ast}_{i}a_{\rm Li}-g^{(R)\ast}_{i}a_{\rm Ri}\right]},\qquad\\  \label{eq:A6}
 \dot{b}&=&-i\omega_{m} b+i\chi a^{\dagger}_{2}a_{1}-i\sum_{i}{g^{\ast}_{\rm bi}b_{i}},\\  \label{eq:A7}
\dot{a_{\rm Li}}&=&-i(\omega_{\rm Li}-\omega_{d})a_{\rm Li}-ig^{(L)}_{i}(a_{1}+a_{2})/\sqrt{2},\\  \label{eq:A8}
\dot{a_{\rm Ri}}&=&-i(\omega_{\rm Ri}-\omega_{d})a_{\rm Ri}-ig^{(R)}_{i}(a_{1}-a_{2})/\sqrt{2}.  \label{eq:A9}
\end{eqnarray}

Under the Markovian approximation, we introduce the decay rates
$\gamma_{L}$, $\gamma_{R}$, and $\gamma_{m}$ of the optical modes
in the left and right cavities, the mechanical mode
as well as the corresponding fluctuation operators~\cite{scully},
then we can obtain the dynamical equation of the cavity modes and
the mechanical modes in Eqs.~(\ref{eq:3})-(\ref{eq:5}) by
noting that $\gamma_{L}=\gamma_{R}=\gamma_{c}$. The  equations of motion
of $J_{-}$, $J_{z}$, and $b$ are obtained in Eqs.~(\ref{eq:14})-(\ref{eq:16}).
With the condition $\gamma_{c}\gg \gamma_{m}$,
by eliminating the operator of the cavity modes with adiabatic approximation,
 we can obtain the equations of motion of the phonon given in Eq.~(\ref{eq:20}),
  with the corresponding fluctuation operator defined as
\begin{eqnarray}\label{eq:A9-1}
\Gamma(t)&=&\alpha\left[\Gamma^{\dagger}_{2}(t)a_{1}
+a^{\dagger}_{2}\Gamma_{1}(t)\right]\nonumber\\
&-&\frac{i\alpha\chi}{\sqrt{2}}\left[\frac{ \Omega\Gamma^{\dagger}_{1}(t)}{N+i\gamma_{c}\Delta }
-\frac{\Omega^{\ast}\Gamma_{2}(t)}{N-i\gamma_{c}\Delta}\right]b\nonumber\\
&+ &\frac{\alpha\left[\frac{\gamma_{c}}{2}-i\left(g-\Delta\right)\right]\Omega\Gamma^{\dagger}_{2}(t)}{\sqrt{2}\left[N+i\gamma_{c}\Delta \right]}+\sqrt{2\gamma_{m}}b_{\rm in}(t).\quad
\end{eqnarray}

\section{Parameters given in subsection~\ref{3c}}\label{A2}

By adiabatically eliminating the cavity variables, the photon inversion operator $J_{z}$ can be written as
\begin{equation}\label{eq:B1}
J_{z}\approx
\frac{\Sigma_{z}(\Delta,\delta)}{(|\varepsilon_{1}|^{2}+4\chi^2
b^{\dagger}b)(N^2+\gamma^2_{c}\Delta^2)}.
\end{equation}
where the term $\Sigma_{z}(\Delta,\delta)$ is defined as
\begin{eqnarray}\label{eq:B2}
\Sigma_{z}&=& g|\Omega|^2\Delta
|\varepsilon_{1}|^{2}+2\chi^2\delta\Delta|\Omega|^2
b^{\dagger}b\\
&-&i\chi|\varepsilon_{1}|^{2}|\Omega|^2\left(b^\dagger
N-N b\right)/\left(2\gamma_{c}\right)\nonumber\\
&+ &\chi|\Omega|^2\left\{i \varepsilon_{1}
\left[(N+\Delta^2)/2+igM/\gamma_{c}\right]b+\text{h.c.}\right\}.\nonumber
\end{eqnarray}
We have defined $\varepsilon_{1}=\gamma_{c}+i\delta$. In the regime near threshold,  $n\ll
|\varepsilon_{1}|^{2}/(4\chi^{2})$, up to the third-order terms
$J_{z}$ can be expanded as follow
\begin{eqnarray}\label{eq:B3}
J_{z}=j_{0}+j_{1}b+j^{\ast}_{1}b^{\dagger}-j_{2}b^{\dagger}b+j_{3}b^{\dagger}b^{\dagger}b+j^{\ast}_{3}b^{\dagger}bb,
\end{eqnarray}
with $j_{0}=g\Delta|\Omega|^{2}\varepsilon_{3}$ and other coefficients $j_{i}$ (with $i=1, 2, 3$)
\begin{eqnarray}
j_{1}&=&i\chi\varepsilon_{3} |\Omega|^{2}\left[M +\left(\varepsilon_{2}M+\gamma_{c}\varepsilon_{1}\Delta^{2}\right)/|\varepsilon_{1}|^{2}\right]/\left(2\gamma_{c}\right),\quad\\ \label{eq:B5}
j_{2}&=&2\chi^{2}\Delta\varepsilon_{3}|\Omega|^{2}\left[\omega_{m}/|\varepsilon_{1}|^{2}
+gM\varepsilon_{3}\right],\quad\\ \label{eq:B6}
j_{3}&=&\left[i\chi^{3}\varepsilon_{3}|\Omega|^{2}/\left(2\gamma_{c}|\varepsilon_{1}|^{2}\right)\right]\nonumber\\
& &\times\left[\left(2\varepsilon_{3}M^{2}-1\right)|\varepsilon_{1}|^{2}+2\varepsilon_{3}\left(\varepsilon_{2}^{\ast}M^{2}+ \gamma_{c}\varepsilon^{\ast}_{1}\Delta^{2}M\right)\right.\nonumber\\
& &-\left.\varepsilon_{2}^{\ast}+4M+4\left(\varepsilon_{2}^{\ast}M+\gamma_{c}\varepsilon^{\ast}_{1}\Delta^{2}\right)/|\varepsilon_{1}|^{2}\right]. \label{eq:B7}
\end{eqnarray}
where the parameters $M=\gamma^{2}_{c}/4+g^2$,
$\varepsilon_{2}=\gamma^{2}_{c}-2g\delta+2ig\gamma_{c}+i\gamma_{c}\delta$, and $\varepsilon_{3}=1/\left(M^{2}+\gamma^{2}_{c}\Delta^{2}\right)$.

Substituting $J_{z}$ in Eq.~(\ref{eq:B3}) into the dynamical
equation of $b$ in Eq.~(\ref{eq:22}), if we define coefficients $G_{i}$ (for $i=0,1,\cdot\cdot\cdot,4$)
\begin{eqnarray}
G_{0}&=&\eta_{1}+i\eta_{2}, \label{eq:B8} \\
G_{1}&=&\eta_{3}-\gamma_{m}-i\left(\omega_{m}+\eta_{4}\right),\\ \label{eq:B9}
G_{2}&=&2\chi^2j^{\ast}_{1}/\varepsilon_{1}+\eta_{5}+i\eta_{6},\\ \label{eq:B10}
G_{3}&=&2\chi^2j_{1}/\varepsilon_{1},\\ \label{eq:B11}
G_{4}&=&\eta_{7}-i\eta_{8}, \label{eq:B12}
\end{eqnarray}
then phonon lasing equation near the threshold can be written as in Eq.~(\ref{eq:24}) with the parameters
\begin{eqnarray}
\eta_{1}&=&\chi \gamma_{c}\varepsilon_{3}|\Omega|^{2}\left(\varepsilon_{4}M+2\delta\Delta^{2}\right)/\left(2|\varepsilon_{1}|^{2}\right),\label{eq:B13}\\
\eta_{2}&=&\chi\varepsilon_{3}|\Omega|^{2}\left(\varepsilon_{5}M+2\gamma^{2}_{c}\Delta^{2}\right)/\left(2|\varepsilon_{1}|^{2}\right),\label{eq:B14}\\
\eta_{3}&=&\gamma_{c}\chi^2\left(2j_{0}-\delta\Delta|\Omega|^{2}\varepsilon_{3}\right)/|\varepsilon_{1}|^{2},\label{eq:B15}\\
 \eta_{4}&=&\chi^2\left(2 \delta
j_{0} +\gamma^{2}_{c}\Delta|\Omega|^{2} \varepsilon_{3}\right)/|\varepsilon_{1}|^{2},\label{eq:B16}\\
\eta_{5}&=&\gamma_{c}\chi^3\varepsilon_{3}|\Omega|^{2}/\left(2|\varepsilon_{1}|^{2}\right)\nonumber\\
& &\times\left[\varepsilon_{4}-2\varepsilon_{3}M|\Omega|^{2}\left(\varepsilon_{4}M+2\delta\Delta^{2}\right)\right],\label{eq:B17}\\
\eta_{6}&=&\chi^3\varepsilon_{3}|\Omega|^{2}/\left(2|\varepsilon_{1}|^{2}\right)\nonumber\\
& &\times\left[\varepsilon_{5}-2\varepsilon_{3}M|\Omega|^{2}\left(\varepsilon_{5}M+2\gamma^{2}_{c}\Delta^{2}\right)\right],\qquad\label{eq:B18}\\
\eta_{7}&=&2\gamma_{c}\chi^2\left(j_{2}-2\delta\chi^{2}\Delta\varepsilon^{2}_{3}|\Omega|^{2}M
\right)/|\varepsilon_{1}|^{2},\label{eq:B19}\\
\eta_{8}&=&2\chi^2\left(\delta
j_{2}+\chi^{2}\gamma^{2}_{c}\Delta\varepsilon^{2}_{3}|\Omega|^{2}M
\right)/|\varepsilon_{1}|^{2}. \label{eq:B20}
\end{eqnarray}
Here we have defined $\varepsilon_{4}=\delta+2g$ and $\varepsilon_{5}=\gamma^2_c-2\delta g$.

Using the semi-classical approximation, the phonon field $b$ can be
written as a two-dimensional vector, $b=(u_{1},u_{2})^T$ .
  Thus, we can obtain the dynamical equations for $u_{1}$ and $u_{2}$ as
\begin{eqnarray}
\dot{ u_{1}}
&=&\eta_{1}+\alpha^{\prime}u_{1}-\text{Im}(G_{1})u_{2}+\left(\eta_{5}+\varepsilon_{6}\right)u^{2}_{1}+\varepsilon_{7}u_{1}u_{2}\nonumber\\
& + &\left(\eta_{5}-\delta\varepsilon_{8}/\gamma_{c}\right)u^{2}_{2}-(u^{2}_{1}+u^{2}_{2})\left(\eta_{7}u_{1}+\eta_{8}u_{2}\right),\label{eq:B21}\\
\dot{ u_{2}}
&=&\eta_{2}+\alpha^{\prime} u_{2}+\text{Im}(G_{1})u_{1}+\left(\eta_{6}-\varepsilon_{8}\right)u^{2}_{2}+\varepsilon_{9}u_{1}u_{2}\nonumber\\
&+ &\left(\eta_{6}-\delta\varepsilon_{6}/\gamma_{c}\right) u^{2}_{1}-(u^{2}_{1}+u^{2}_{2})\left(\eta_{7}u_{2}-\eta_{8}u_{1}\right). \label{eq:B22}
\end{eqnarray}
with the corresponding coefficients
\begin{eqnarray}
\varepsilon_{6}&=&4\chi^2  \gamma_{c}\text{Re}(j_{1})/|\varepsilon_{1}|^{2},\label{eq:B23} \\
\varepsilon_{7}&=&4\chi^{2}\left[\delta\text{Re}(j_{1})-\gamma_{c} \text{Im}(j_{1})\right]/|\varepsilon_{1}|^{2},\label{eq:B24}\\
\varepsilon_{8}&=&4\chi^2 \gamma_{c}\text{Im}(j_{1})/|\varepsilon_{1}|^{2},\label{eq:B25}\\
\varepsilon_{9}&=&4\chi^{2}\left[\gamma_{c}\text{Re}(j_{1})+\delta\text{Im}(j_{1})\right]/|\varepsilon_{1}|^{2}.\label{eq:B26}
\end{eqnarray}

In the special case of $\delta=0$, then the conditions $\varepsilon_{6}\gg
\varepsilon_{8},\eta_{5},\eta_{6}$, and also $\eta_{7}\gg \eta_{8}$,
 the dynamical equations of $u_{1}$ and $u_{2}$ can be written as
\begin{eqnarray}
\dot{ u_{1}}
&=&\eta_{1}+\alpha^{\prime}u_{1}-\text{Im}(G_{1})
u_{2}+\varepsilon_{6}u^{2}_{1}-\varepsilon_{8}u_{1}u_{2}+\eta_{5}u^{2}_{2}\nonumber\\
& &+\eta_{8}(u^{2}_{1}+u^{2}_{2})u_{1}-\eta_{8}(u^{2}_{1}+u^{2}_{2})u_{2}.\label{eq:B27}  \\
\dot{ u_{2}}
&=&\eta_{2}+\text{Im}(G_{1})u_{1}+\alpha^{\prime} u_{2}
+\varepsilon_{6}u_{1}u_{2}+\left(\eta_{6}-\varepsilon_{8}\right)u^{2}_{2}\nonumber\\
& &+\eta_{6}u^{2}_{1}+\eta_{8}(u^{2}_{1}+u^{2}_{2})u_{2}+\eta_{8}(u^{2}_{1}+u^{2}_{2})u_{1}. \label{eq:B28}
\end{eqnarray}
The conditions $\varepsilon_{7}=-\varepsilon_{8}$ and $\varepsilon_{9}=\varepsilon_{6}$
(for $\delta=0$) have been used to obtain these equations.

\section{Derivation for the dynamical equations for the fluctuation operators}\label{A3}

Under the small fluctuation approximation as shown in
Eqs.~(\ref{eq:10})-(\ref{eq:12})
and from Eqs.~(\ref{eq:3})-(\ref{eq:5}), we can write down
the dynamical equations for the fluctuation operators as
\begin{eqnarray}
\dot{\Lambda}_{1}(t)&=&-\left[\gamma_{c}/2+i(g-\Delta)\right]\Lambda_{1}(t)+\Gamma_{1}(t)\nonumber\\
& &+i\chi\left[A_{2}\beta(t)+B_{0}\Lambda_{2}(t)\right],\label{eq:C1}\\
\dot{\Lambda}_{2}(t)&=&-\left[\gamma_{c}/2-i(g+\Delta)\right]\Lambda_{2}(t)+\Gamma_{2}(t)\nonumber\\
& &+i\chi\left[A_{1}\beta^{\dagger}(t)+B^{\ast}_{0}\Lambda_{1}(t)\right], \label{eq:C2}\\
 \dot{\beta}(t)&=&-\left(\gamma_{m}+i\omega_{m} \right)\beta(t)+\sqrt{2\gamma_{m}}b_{\rm in}(t)\nonumber\\
 & &+ i\chi\left[A^{\ast}_{2}\Lambda_{1}(t)+A_{1}\Lambda^{\dagger}_{2}(t)\right].\label{eq:C3}
\end{eqnarray}
If we introduce the Fourier transform
$f(t)=\int^{+\infty}_{-\infty}{f(\omega)\exp{(-i\omega
t)}(d \omega/2\pi)}$ for arbitrary smooth function $f(t)$, the motion equations for the
fluctuation operators in the frequency domain can be written as
\begin{eqnarray}
-i\omega\tilde{\Lambda}_{1}(\omega)&=&-\left[\gamma_{c}/2+i(g-\Delta)\right]\tilde{\Lambda}_{1}(\omega)+\tilde{\Gamma}_{1}(\omega)\nonumber\\
& &+i\chi\left[A_{2}\tilde{\beta}(\omega)+B_{0}\tilde{\Lambda}_{2}(\omega)\right],\label{eq:C4}\\
-i\omega\tilde{\Lambda}_{2}(\omega) &=&-\left[\gamma_{c}/2-i(g+\Delta)\right]\tilde{\Lambda}_{2}(\omega)+\tilde{\Gamma}_{2}(\omega)\nonumber\\
& &+i\chi \left[A_{1}\tilde{\beta}^{\dagger}(\omega)+B^{\ast}_{0}\tilde{\Lambda}_{1}(\omega)\right], \label{eq:C5} \\
-i\omega\tilde{\beta}(\omega) &=&-\left(\gamma_{m}+i\omega_{m}\right) \tilde{\beta}(\omega)+\sqrt{2\gamma_{m}}\tilde{b}_{\rm in}(\omega)\nonumber\\
& &+i\chi\left[A^{\ast}_{2}\tilde{\Lambda}_{1}(\omega)+A_{1}\tilde{\Lambda}^{\dagger}_{2}(\omega)\right]. \label{eq:C6}
\end{eqnarray}

By eliminating the fluctuation operators
$\tilde{\Lambda}_{1}(\omega)$ and $\tilde{\Lambda}_{2}(\omega)$,
we can obtain the expression of $\tilde{\beta}(\omega)$ in Eq.~(\ref{eq:59}). The corresponding
coefficients $p_{i}$ $(i=1,2,\cdot \cdot\cdot, 6)$ are given by
\begin{eqnarray}
p_{1}&=&\left(\sqrt{2\gamma_{m}}\lambda_{1}-\chi^{2}A^{\ast}_{2} B_{0} n_{2}\right)/D_{1},\label{eq:C7}\\
p_{2}&=&i\chi\left( A_{1}n^{+}_{2}\lambda_{1}+i\chi A^{\ast}_{2}B_{0} n_{1} m_{2} \right)/D_{1},\label{eq:C8}\\
p_{3}&=&i\chi\left[ A_{1}n^{+}_{3}\lambda_{1}+ A^{\ast}_{2}+i\chi A^{\ast}_{2} B_{0} (n_{4}+n_{1}m_{3})\right]/D_{1},\qquad\label{eq:C9}\\
p_{4}&=&i\chi\left[ A_{1}n^{+}_{4}\lambda_{1}+i\chi A^{\ast}_{2}B_{0} (n_{3}+n_{1}m_{4})\right]/D_{1},\label{eq:C10}\\
p_{5}&=&i\chi\left[ A_{1}n^{+}_{5}\lambda_{1}+i\chi
A^{\ast}_{2}B_{0} (n_{6}+n_{1}m_{5})\right]/D_{1},\label{eq:C11}\\
p_{6}&=&i\chi\left[ A_{1}n^{+}_{6}\lambda_{1}+i\chi
A^{\ast}_{2}B_{0} (n_{5}+n_{1}m_{6})\right]/D_{1},\label{eq:C12}
\end{eqnarray}
and the corresponding coefficients $\lambda_{i}(\omega)$ ($i=1,2,\cdot \cdot\cdot, 4$),
$m_{i}(\omega)$ ($i=1,2,\cdot \cdot\cdot, 6$), and $n_{i}(\omega)$ ($i=1,2,\cdot \cdot\cdot, 6$)
are defined as
\begin{eqnarray}
\lambda_{1}&=&\gamma_{c}/2+i\left(g-\Delta\right)-i\omega,\\
\lambda_{2}&=&\gamma_{c}/2-i\left(g+\Delta\right)-i\omega,\\
\lambda_{3}&=&\gamma_{m}+i\omega_{m}-i\omega,\\
\lambda_{4}&=&\lambda_{1}\lambda_{2}+ \chi^{2}|B_{0}|^{2},\\
m_{1}&=&i\chi^{3}A^{\ast}_{1}A_{2}B^{\ast}_{0}\left(\lambda_{4}+\lambda^{+}_{4}\right)/\left(D_{2}\lambda^{+}_{3}\lambda_{4}\lambda^{+}_{4}\right),\qquad\\
m_{2}&=&\sqrt{2\gamma_{m}}/\left(\lambda^{+}_{3}D_{2}\right),\quad\\
m_{3}&=&\chi^{2} A^{\ast}_{1}B^{\ast}_{0}/\left(\lambda^{+}_{3}\lambda_{4}D_{2}\right), \\
m_{4}&=& i\chi A_{2}\left(\chi^{2} |B_{0}|^{2}-\lambda^{+}_{4}\right)/\left( \lambda^{+}_{1}\lambda^{+}_{3}\lambda^{+}_{4}D_{2}\right),\\
m_{5}&=&-i\chi\lambda_{1}A^{\ast}_{1}/\left(\lambda^{+}_{3}\lambda_{4}D_{2}\right),\\
m_{6}&=&-\chi^{2} A_{2}B^{\ast}_{0}/\left(D_{2}\lambda^{+}_{3}\lambda^{+}_{4}\right),\\
n_{1}&=& i\chi \left( A_{1}\lambda_{1}+i\chi
B^{\ast}_{0}A_{2}m^{+}_{1}\right)/ \lambda_{4},\\
n_{2}&=&-\chi^{2}B^{\ast}_{0} A_{2}m^{+}_{2}/ \lambda_{4},\quad\\
n_{3}&=&- \chi^{2}B^{\ast}_{0}A_{2}m^{+}_{3}/\lambda_{4},\\
n_{4}&=&  i\chi\left( B^{\ast}_{0}+ i\chi B^{\ast}_{0} A_{2}m^{+}_{4}\right)/\lambda_{4}, \\
n_{5}&=&-\chi^{2}B^{\ast}_{0}A_{2}m^{+}_{5}/\lambda_{4},\\
n_{6}&=&\left(\lambda_{1}-\chi^{2}B^{\ast}_{0}A_{2}m^{+}_{6}\right)/\lambda_{4}.
\end{eqnarray}
with
\begin{equation}
D_{1}=\lambda_{1}\lambda_{3}-i\chi A_{1}n^{+}_{1}\lambda_{1}+\chi^{2}A^{\ast}_{2}
B_{0}n_{1}m_{1}+\chi^{2} |A_{2}|^{2},
\end{equation}
and $ D_{2}=1+\chi^{2}|A_{2}|^{2}/\left(\lambda^{+}_{1}\lambda^{+}_{3}\right)-\chi^{2}|A_{1}|^{2}\lambda_{1}/\left( \lambda^{+}_{3}\lambda_{4}\right)
-\chi^{4}|B_{0}|^{2}|A_{2}|^{2}/\left(\lambda^{+}_{1}\lambda^{+}_{3}\lambda^{+}_{4}\right).$

\end{document}